\documentclass[journal,twoside,web]{ieeecolor}

\usepackage{generic}
\usepackage{cite}

\usepackage{amsmath,amssymb,amsfonts}
\usepackage{amsthm}

\usepackage{graphicx}
\graphicspath{{figure/}}

\usepackage[table,xcdraw]{xcolor}

\usepackage[hidelinks]{hyperref}
\usepackage{url}

\usepackage{comment}
\usepackage{multirow}
\usepackage{array}
\usepackage{overpic}

\usepackage{subcaption}

\usepackage{algorithm}
\usepackage{algpseudocode}
\algrenewcommand\algorithmicrequire{\textbf{Input:}}
\algrenewcommand\algorithmicensure{\textbf{Initialize:}}

\theoremstyle{remark}

\allowdisplaybreaks[1]

\begin{document}
\title{{{Cyber Security of Sensor Systems for State Sequence Estimation: A   Machine Learning   Approach }
} 
	\thanks{This work was sponsored by the Office of Naval Research (ONR) under grant number N00014-22-1-2626. }
}

\author{ Xubin Fang, Rick S. Blum, \IEEEmembership{Life Fellow, IEEE}  
\thanks{Xubin Fang and Rick S. Blum
are with the Department of Electrical and Computer Engineering, Lehigh University, Bethlehem, PA 18015 USA (e-mails: rblum@eecs.lehigh.edu, 
xuf220@lehigh.edu 
).}, Ramesh Bharadwaj\thanks{Ramesh Bharadwaj is with the Naval  Research Laboratory,  Wahington DC USA (email: ramesh.bharadwaj@nrl.navy.mil).}, and Brian M. Sadler \IEEEmembership{Life Fellow, IEEE}%
	\thanks{Brian M. Sadler is with UT-Austin, Austin, TX USA (e-mail: Brian.sadler@ieee.org).} }
\maketitle

\begin{abstract}
Due to possible devastating consequences, counteracting sensor data attacks is an extremely important topic, which has not seen sufficient study. 
{To the best of our knowledge, this paper develops the first methods 
that accurately identify/eliminate only the problematic attacked sensor data presented to a sequence estimation/regression algorithm under any attack from our attack model. }
The approach does not assume a known form for the statistical model of the sensor data, allowing data-driven and machine learning sequence estimation/regression algorithms to be protected. 
A simple protection approach for attackers not endowed with knowledge of the details of our protection approach is first developed, followed by  additional processing for attacks based on protection system knowledge. 
Experimental results show that the simple approach achieves  performance indistinguishable
from that for an approach  which knows which sensors are attacked. 
For cases where the attacker has knowledge of the protection approach, experimental results
indicate the additional processing can be configured so that the worst-case degradation under the additional processing and a large number of sensors attacked  can be made significantly smaller than 
the worst-case degradation of the simple approach, and close to an approach which knows which sensors are attacked,  
with just a slight degradation under no attacks. 
Mathematical descriptions of the worst-case attacks are used to demonstrate 
the additional processing will provide similar advantages for cases for which we do not have numerical results. 
All the data-driven/machine learning  processing used in our approaches employ only unattacked training data.  
\end{abstract}


\section{Introduction}
Incorporation of sensors into infrastructure  provides important advantages 
\cite{Varshney,Willett,BChen,RNiu,RVish,LKaplan,Visa}. 
The 2020 World Economic Forum’s Global Risks Report listed cyber attacks on global critical infrastructure as a top concern that urgently needs to be mitigated \cite{GRR2020}.  
Sensors are highly vulnerable to cyber attacks and cyber attacks on sensors can cause tremendous damage. 
Unfortunately, protection against such cyber attacks on sensors has not been adequately addressed \cite{sensors-security}. 
Here we focus on protecting sensor-based sequence  estimation/regression algorithms.
 These  
sensor-based sequence estimation/regression algorithms can be employed to estimate  any quantity sensed by the sensors, for example: position, velocity and acceleration of objects of interest.  

Many systems today depend on sensors.  These include  vehicles and vehicle networks, internet of things systems, 
and other smart systems. There are attacks on sensors that 1) give the attacker complete control over the attacked data; 
2) allow attacks at a sensor to be changed arbitrarily vs time; 3) allow attacked sensors to be changed arbitrarily vs time; 4) Allow the number of sensors attacked to be changed arbitrarily vs time; 5) provide no prior knowledge on which sensors are more likely attacked. Mitigation approaches for such cases are lacking so we consider them here.

\begin{figure}
\centering
\begin{subfigure}[b]{0.55\textwidth}
   \includegraphics[width=.8\linewidth,height=90pt]{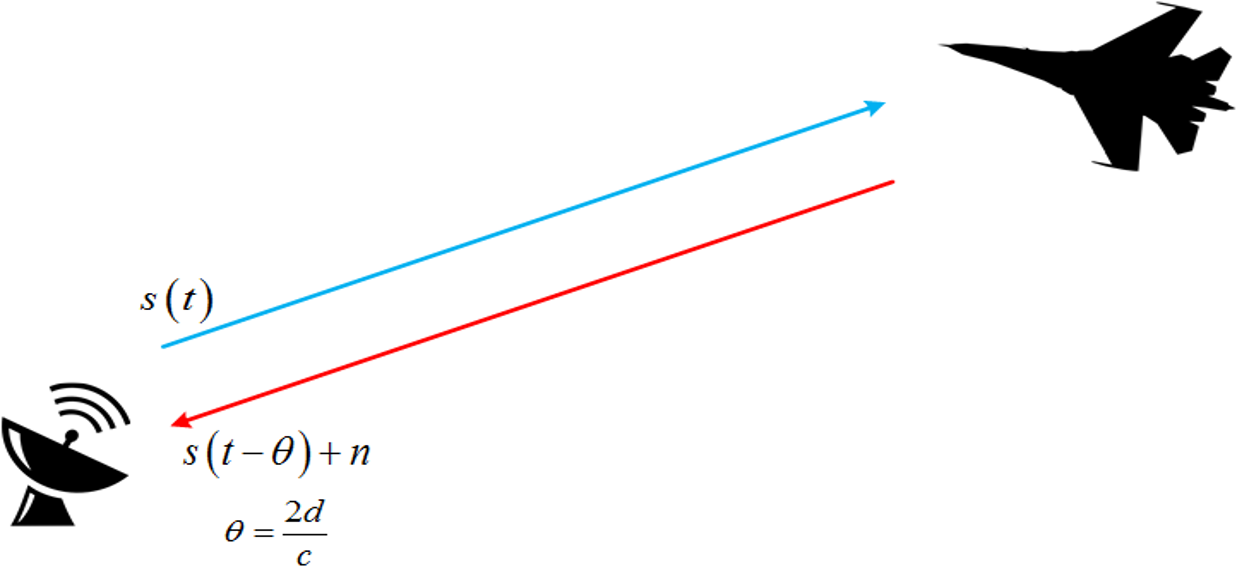}
   \caption{}
   \label{fig:Ng1} 
\end{subfigure}
\begin{subfigure}[b]{0.55\textwidth}
   \includegraphics[width=.8\linewidth,height=90pt]{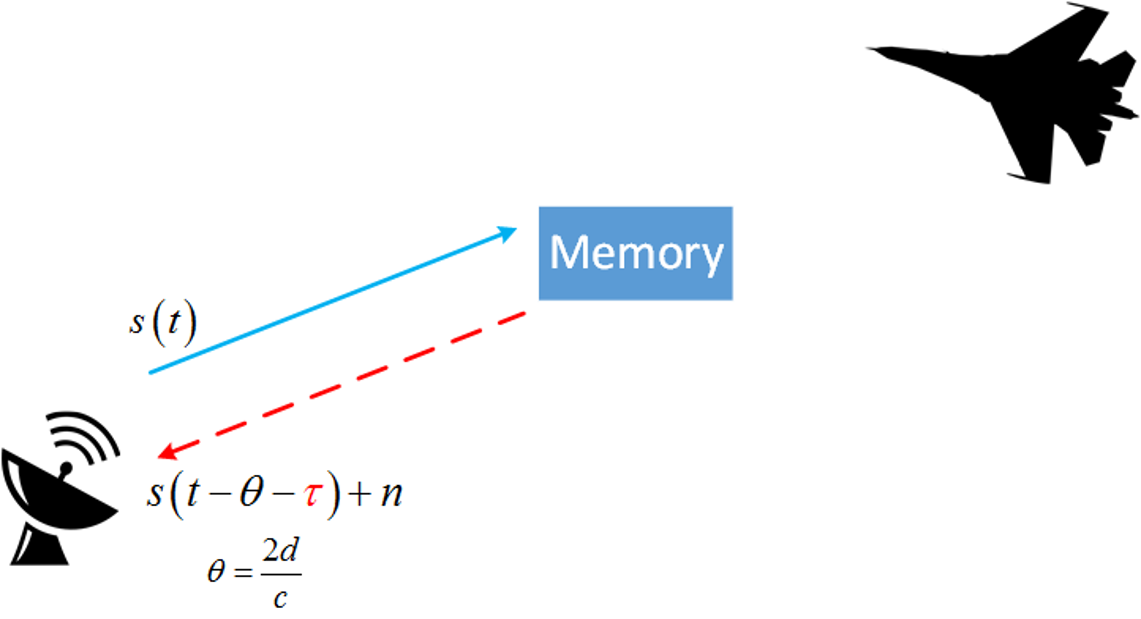}
   \caption{}
   \label{fig:Ng2}
\end{subfigure}
\caption[Figure illustrating spoofing attack on a radar.]{ Figure illustrating spoofing attack on a radar. (a) Unattacked. (b) Attacked.}
\label{fradar} \end{figure}

Consider a radar system as one example of an important sensor that can be attacked. During unattacked operation, as shown in Fig~\ref{fradar}a, a radar transmits a pulsed wave in a given direction, which after bouncing off an object, maybe an airplane, is reflected back towards the radar.  The pulsed wave is received at the radar with a given delay $\theta$ with respect to the transmitted waveform and with additive noise and clutter $n$. By knowing the speed of the wave $c$, the distance $d$ from the radar to the object can be determined based on the delay.  As shown in 
Fig~\ref{fradar}b, 
a spoofing attack can be launched if an attacker receives the waveform  transmitted by the radar,  stores it in a digital memory for a while, and then plays back the waveform with an extra delay $\tau$ of its choosing.  Since the attacker can choose the extra delay, he has complete control of the distance to the object that the radar reports, the sensor data.   

Similar attacks apply to other active sensor systems\footnote{Active sensor systems transmit signals.}, including lidar, sonar, 
and GPS. In GPS attacks, attackers often transmit fake GPS  signals \cite{GPS}.  All such attacks are sometimes called in-band attacks since the attacks employ signals whose frequencies match those sensed by the sensor. 
Many other sensors allow similar attacks, including  many of the in-band and out-of-band attacks described in 
\cite{oob}. These  out-of-band attacks employ signals of vastly different frequencies from those sensed,  impinging on sensors or connections to alter sensor data.  The impinging attack signal modality can be acoustic, optical, or electromagnetic, while the sensor senses a different modality.  One group of out-of-band attacks employ acoustic attacks targeting the resonant frequencies of gyroscopes and accelerometers.  
Many of these attacks, along with many  classical spoofing and man-in-the-middle attacks, which modify transmissions of sensor data   \cite{addOurSPmag}, 
allow attackers to replace actual sensor data with values of their choosing, a powerful type of attack we consider later. 

\subsection{Problem and Attack Model} \label{ss:Pandattack} 


This work aims to provide an outer shell of protection that can surround any unprotected sequence estimation/regression algorithm chosen from a large class.  
Without the shell, the original unprotected estimation algorithm is designed for cases without cyber attacks on the sensor data. The shell will eliminate problematic (in terms of degrading the estimate) attacked sensor data, allowing the protected estimation algorithm to operate using the remaining data. 
We focus on centralized sequence 
estimation algorithms in this paper, but we intend to later consider decentralized estimation algorithms in an attempt to show similar ideas can be employed there. 

The attack model considered in this paper allows much more powerful attacks than we have ever seen successfully detected/eliminated  in the existing attack mitigation literature. We assume the attacker has complete control to generate any sensor data values after the attack.  We assume the protection system has no prior knowledge of which sensors are more likely attacked.  We assume the attacked sensors and the attacks launched on those sensors  can change each time step.   We assume the protection system can't make any assumptions on the number of sensors attacked at any time step or about the patterns of attacks over time. Such attacks are consistent with sensor spoofing and out-of-band attacks discussed in Subsection~\ref{ss:Pandattack}, where the signal impinging on the sensor is already modified and this modification changes a sensor output exactly as desired by the attacker.  The attacked sensors can also exhibit such properties for other types of sensor attacks also, see Subsection~\ref{ss:Pandattack} . We provide two protection approaches, one for attackers without detailed knowledge of the protection approach and another for attackers with detailed knowledge of our protection approach. We assume all the data-driven/machine learning  processing used in our approaches, for protection or for estimation, will employ only unattacked training data since it seems impossible to obtain training data accurately representing all possible attacks. 

\subsection{Contribution} 

{To the best of our knowledge, this paper develops the first methods that accurately 
identify/eliminate all problematic attacked sensor data (keep all the rest) 
presented to a sequence   estimation/regression algorithm 
for all possible attacks in our 
attack model. 
As we discuss later in the literature review, there are other papers which make assumptions on things like the number of sensors attacked and when these assumptions are not true (yielding attacks in our attack model), these approaches provide very poor performance at the just described task.  
Our approaches can be used as an outer shell of protection that can surround an unprotected estimation/regression algorithm (designed for unattacked sensor data), 
allowing the protected estimation 
algorithm to properly operate, with excellent performance, using the deemed to be unattacked data for 
any attack in our attack model.  
Our approaches do not assume a known form for the statistical model of the sensor data, allowing data-driven and machine learning sequence estimation/regression algorithms to be protected.  
Our approaches employ  only unattacked training data 
and handle the possible attacks allowed by our powerful attack model 
where  the attacked sensors and attacks can change for each time step, the attacker has complete control of after-attack sensor data, and the protection system has no prior knowledge of which sensors are more likely attacked and how many sensors are attacked. 
}
 
We initially focus in this paper on a simple protection approach for attacks not endowed with knowledge of the details of our protection approach, but later we describe additional processing for attackers with knowledge, thus allowing a lower complexity approach if suitable. 
{The paper provides experimental demonstration of good performance
for both the simple and additional processing. 
Our simple method is shown to achieve nearly 
indistinguishable estimation performance in cases tested, which assume the attacker does not have knowledge of the protection scheme, 
in comparison to 
an optimized approach which knows which sensors were attacked.  
For cases where the attacker has knowledge of the protection approach, the additional processing can be configured so that the worst-case degradation under the additional processing and a large number of sensors attacked  can be made significantly smaller than 
the worst-case degradation 
under the simple approach, and close to that for an approach which knows which sensors are attacked,  
for the same number of attacked  sensors with just a slight degradation under no attacks.} 
Guarding against the worst-case is extremely important and if this performance can be made acceptable, then performance will always be acceptable for any attacks. 

To accomplish the majority of what we have just described, we rely heavily on our discovery on how to mathematically describe the worst-case attacks for our simple and additional processing approaches. This then allowed us to describe how to calculate the worst-case performance. 
We feel this is an 
important contribution of our work that can be employed to analyze other protection approaches in a similar manner.  We hope others will employ these ideas in future work. While our numerical testing is limited, as all would be, the mathematical description of our worst-case attacks allow us to ensure, even for cases not tested, that proper choice of the additional processing parameters imply 
the worst-case degradation under the additional processing and a large number of sensors attacked  can be made significantly smaller than 
the worst-case degradation for the 
simple approach for the same number of sensors attacked. 
Directly calculating the worst-case performance, by knowing the worst-case attack, is extremely efficient in reducing computations, which is especially important if you want to try many parameter settings. It avoids trying many different attacks, an approach others take. 
This would be difficult to use due to extremely high complexity and one would never obtain the actual worst-case attack performance. 


\subsection{Example Application}

To provide a specific example application, we consider connected vehicle networks (CVNs). However the general ideas can be applied to other 
applications. 
In CVNs, sensor technology is being adopted by automotive manufacturers, creating distributed self-organized networks of many high-speed vehicles and infrastructure \cite{z1} where these vehicles can communicate with each other to share sensor data to make driving safer \cite{z2}. 
Using sensor data to help identify the time-varying position and velocity of objects of interest  
is an important fundamental building block in CVNs 
\cite{z3,z5} since it is used by most other required functions, such as navigation and
collision avoidance \cite{z6}.  It is well known that CVNs are vulnerable to sensor attacks \cite{sensorattack,sensorattack2,sensorattack3} 
so this seems a good example application we can use as needed in the rest of the paper.  


Section~\ref{sec:LR} provides a literature review.  In Section~\ref{sec:No_know} we describe our proposed methods and the worst-case attacks on these methods.  Section~\ref{sectionER} presents our numerical results. 
In Section~\ref{sectcon} we provide conclusions. 


\section{Literature Review} \label{sec:LR} 

There has been study on protecting classical sensor-based sequence estimation algorithms, for example Kalman filtering approaches, which assume a known mathematical model (usually linear) for the sensor data, 
see \cite{z18,z19,z20,z21,z22,zwadd} for example. As these protection methods exploit the known mathematical sensor data model, 
they are not applicable to protecting data-driven (machine learning) approaches where a known mathematical sensor data model is not available. 
While these methods are not applicable to the problem of interest here, it is worth noting that the methods of this type that achieve the best performance generally perform a search over all the possible subsets of sensors which could be attacked.  Such searches greatly increase the implementation complexity and run time of these algorithms. Our goal is to avoid such searches to avoid these issues. 

Another approach to the problem of inference with attacked sensor data has been studied under the topic of inference 
 with Byzantine data \cite{Varsh-Byz,{Varsh2}}. 
This work built on early work on the Byzantine Generals problem \cite{byzadd}. 
Unfortunately, these approaches are known to provide very poor performance for some attacks when the number of attacked sensors is unknown, which makes them unsuitable for our attack model.  For example, it is common in these approaches to assume that more than half of the sensors are unattacked and to produce an estimate based on identifying, and then using, these sensors.  This can yield very poor performance if the assumption is not true.  Other assumptions lead to similar issues. 

The Byzantine ideas have also been used in the distributed 
and federated learning paradigm \cite{byzMachLearn}, where processing nodes/agents share data and some nodes/agents launch attacks.  In the distributed 
and federated learning paradigm, the agents may pass gradients as opposed to raw data, but the goal of rejecting the gradients representing attacked data, while keeping the gradients representing unattacked data,  makes the problems similar in some sense.  In the distributed 
and federated learning paradigm, computing the average of the unattacked gradients becomes very important and thus the robust computation of these averages, called robust aggregation, has also become important.  One example of robust aggregation employs a median operation, which is robust to attacks unless the number of values (gradients) attacked is greater than half of the 
number of values aggregated, the number of sensors in our cases.  Trimmed means/medians and related approaches (see \cite{byzMachLearn}), which have also (along with the median) received attention in the signal processing community \cite{orderinSP}, are also applicable. Clearly these robust aggregation approaches are also applicable to our sensor attack problems.  Unfortunately, for the attack model we consider these robust aggregation approaches 
can perform poorly for some attacks. In fact, \cite{byzMachLearn}) states that none of these robust aggregation approaches can be guaranteed to give accurate aggregation results when the number of attacked sensors is unknown.

More recently, some distributed estimation approaches have been considered where the notion of the trust in an agent is available \cite{Goldsmith} in a multi-agent system.   The trust describes prior knowledge, possibly from past interactions, on the likelihood that an agent is providing false or attacked information.  It has been shown that when trust information is available, it is possible to significantly outperform the robust aggregation approaches \cite{trustCOV1}, \cite{trustCOV2}.  In some cases, one can even provide accurate estimates for cases where more than half of the agents are providing false/attacked data \cite{Goldsmith} if trust information is available. 
Unfortunately, the nature of the attacks considered here would not allow one to produce the required trust information. We simply do not have any information on which sensors are more likely to be attacked.  
 

Recently, some machine learning-based encoder-decoder anomaly detection (EDAD) approaches were proposed for protecting general sequence estimation algorithms and significant progress on advancing the technology has taken place, see the full story in \cite{ad-gan-survey1,ad-gan-survey2, ad-gan-survey3}. 
In particular, there have been a rapid series of papers which have presented further improvement of the initial basic EDAD approaches employed.  The approaches have progressed from using early technology, for example autoencoder technology, to much more sophisticated approaches, the latest being generative adversarial network technology.
Using these anomaly detection approaches to protect against sensor attacks has several advantages: (1) No assumptions needed on the number of attacked sensors or amount of attacked sensor data. (2) We can take advantage of the great progress to incorporate the latest technology. (3) There are many available approaches that need only unattacked training data. 
Unfortunately, all of these 
EDAD approaches 
have a negative point which we discuss next.  However, we have a method to augment these approaches that overcomes this negative point.

The EDAD methods learn  a statistical model fitting all the possible (called valid) observed time sequences of unattacked sensor training data.  The  
training data set is assumed to be sufficiently large to fully describe all unattacked sensor data.  During anomaly detection, any sensor data not following that model will be marked as anomalous (either an attacked or broken sensor, we call both attacked here).  
Thus, the EDAD methods will identify data that is not consistent with the training data as anomalous. For example, if an  observed sensor data sequence  exhibits certain patterns in training data but different patterns during operation,  
an anomaly (attack) is detected. 
This enables very powerful verification that the sensed trajectories are possible under no anomaly. 

On the other hand, these EDAD approaches have an issue for the trajectory state estimation problems we consider here which typically have more than one possible (valid) state sequence.  
To make things clear, consider the case  where the state sequence being estimated is the position/velocity of some real object.  There will be many possible (valid) state sequences in such an application but many of them do not follow the position/velocity of the real object we are tracking. 
If the attacker  substitutes sensor data following one valid sequence/trajectory with sensor data following another valid sequence/trajectory, then this attack passes EDAD. This gives the attacker tremendous power to cause big problems.  
{ The attacker can lead the system to believe the state being estimated is following a very different trajectory than it is actually following, potentially inducing an extremely large error in the  estimates based on this sensor data. This error can be as large as the error between two valid trajectories that are farthest apart, unbounded if this difference is unbounded. 
Similar problems occur 
if we estimate the state of some machine or { some other state sequence. }
 
 }


To overcome this issue, after EDAD we employ novel additional checks   
that are shown to provide acceptably  small degradations  even with full knowledge attacks on a large number of sensors. 
The additional checks makes use of a machine learning  predictor trained to predict a typical unattacked sensor's measurement of the actual state trajectory, for example the position/velocity of the real object we are tracking.  
Details are described later.  We have not seen this approach used to fix issues with EDAD.  In fact, we have not seen these issues discussed, even though EDAD has been suggested for state sequence estimation problems. 


\section{Proposed Methods and worst-case Attacks}\label{sec:No_know}

Next, we describe our two approaches for augmenting EDAD to eliminate the issue described at the end of the previous section.  Our approaches implement what we call an actual path consistency check (APCC) to test if the sensor data is following the actual trajectory of the state. 
The first approach, called APCC-SIMPLE, 
is designed for cases where the attacker does not have knowledge about the details of the protection approach.  
 The second approach, called APCC-ADDITIONAL,  is designed for cases when the attacker has knowledge about the protection approach. 
 This second approach adds some additional checks.

\begin{figure*}[h!]
	\centering	
    {\includegraphics[width=0.99\linewidth]{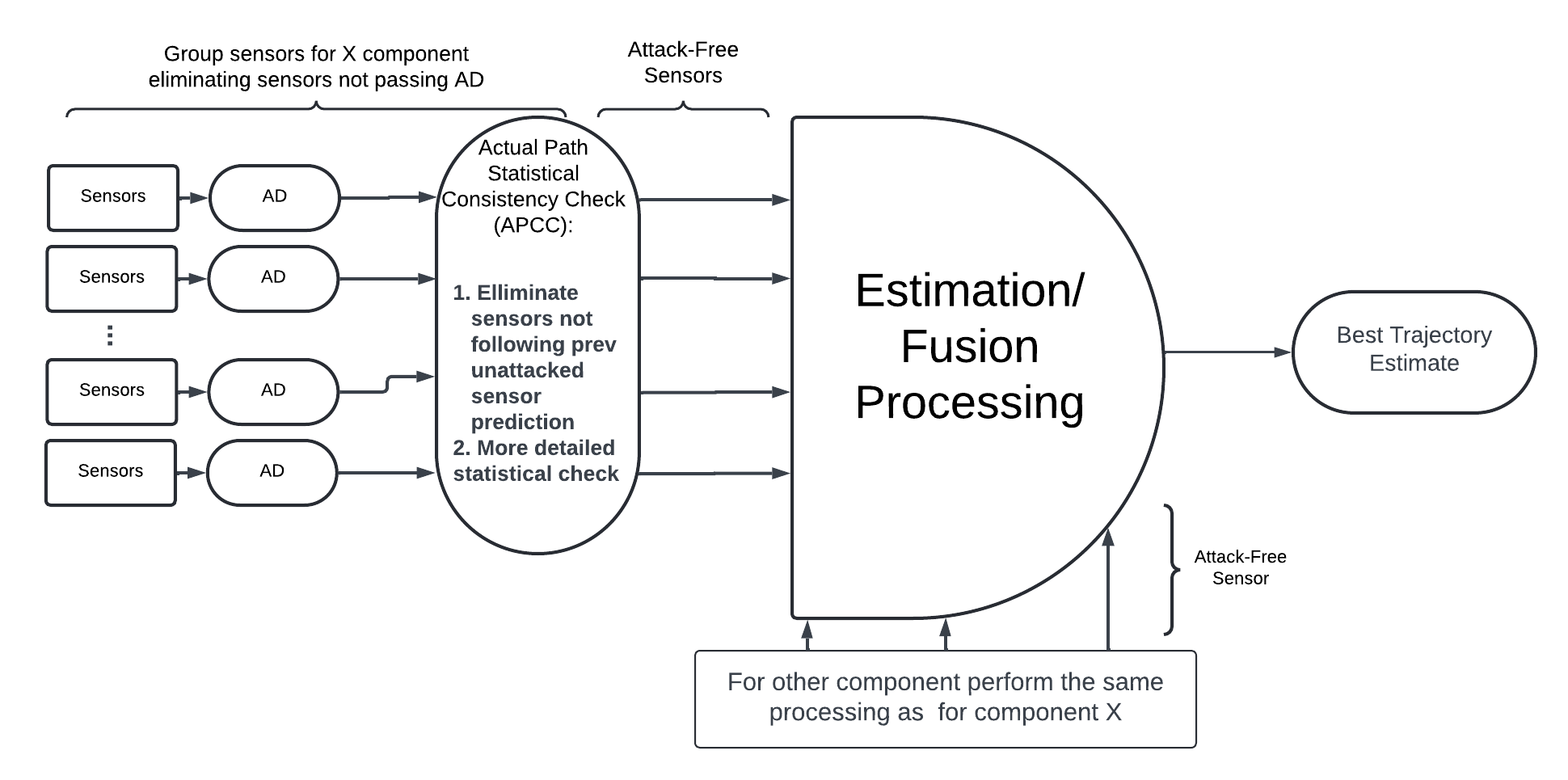 }{
	\caption{{Block diagram of our approach. The block AD is an EDAD approach as per the text. We have two APCC options, APCC-SIMPLE (1. only) and APCC-ADDITIONAL (both 1 and 2) in the APCC block in the figure.}}
	\label{f1}}
    }
\end{figure*}

First, we group all 
similar 
sensors such that each group contains all processed sensor outputs that predict the same component of the tracked trajectory. 
For the group of sensors predicting the first component of the trajectory (maybe the $X$ component of an object's position as in the top left of Fig~\ref{f1}), all the sensors passing EDAD (AD in Fig~\ref{f1}) are put through our APCC, which attempts to check if the sensor data is following the path of the actual trajectory.  This same processing is carried out for each group of sensors predicting all the other components of the trajectory.  Then all the sensor data that passes this check, deemed to be unattacked sensor data, will be sent on to the estimation/fusion processing which will combine (fuse) this data, possibly with other data (including trajectory estimates at previous time steps), to produce the trajectory estimate. The estimation/fusion processing can be thought of as a possibly user-selected estimation algorithm that assumes all the sensor data input to it is unattacked.  

The overall approach is illustrated in the block diagram in Fig~\ref{f1}. It should be noted that the APCC block includes the two options just discussed.  The first option, called APCC-SIMPLE, consists of just the part labeled 1  in the APCC block in Fig~\ref{f1}.  The second option, 
called APCC-ADDITIONAL, 
includes both the parts labeled 1 and 2 (2 builds on 1) in the APCC block in Fig~\ref{f1}. 
Next, we describe the details of the APCC block, under two subsections entitled  APCC-SIMPLE and APCC-ADDITIONAL.  This is followed by a subsection on the worst-case attacks.

\subsection{APCC-SIMPLE}\label{APCCS}
Let us remind the reader that this version of the APCC is employed alone for cases where the attacker has no knowledge of the protection approach. 
For attackers with knowledge, we augment this approach as described in the next subsection. For APCC-SIMPLE, 
take any sensor measurements in a given group that have passed EDAD at a given time step and immediately subtract from them a prediction of what a typical unattacked sensor output would look like at that time step.  
The prediction comes from a trained machine learning algorithm whose input is sensor data from previous time steps which we deemed to be unattacked
and possibly predictions at previous times.  After the subtraction, 
we check that the difference lies in an interval chosen to contain a certain percentage $\beta$ of all the possible values observed when computing this difference using all  the unattacked training data.  We refer to this percentage $\beta$ as the consistency percentage. In some numerical results we select $\beta$ to be $99.9\%$. Then any sensors producing differences not falling within the interval will be deemed  attacked.

\begin{figure}[h!]
	\centering	\includegraphics[width=0.8\linewidth]{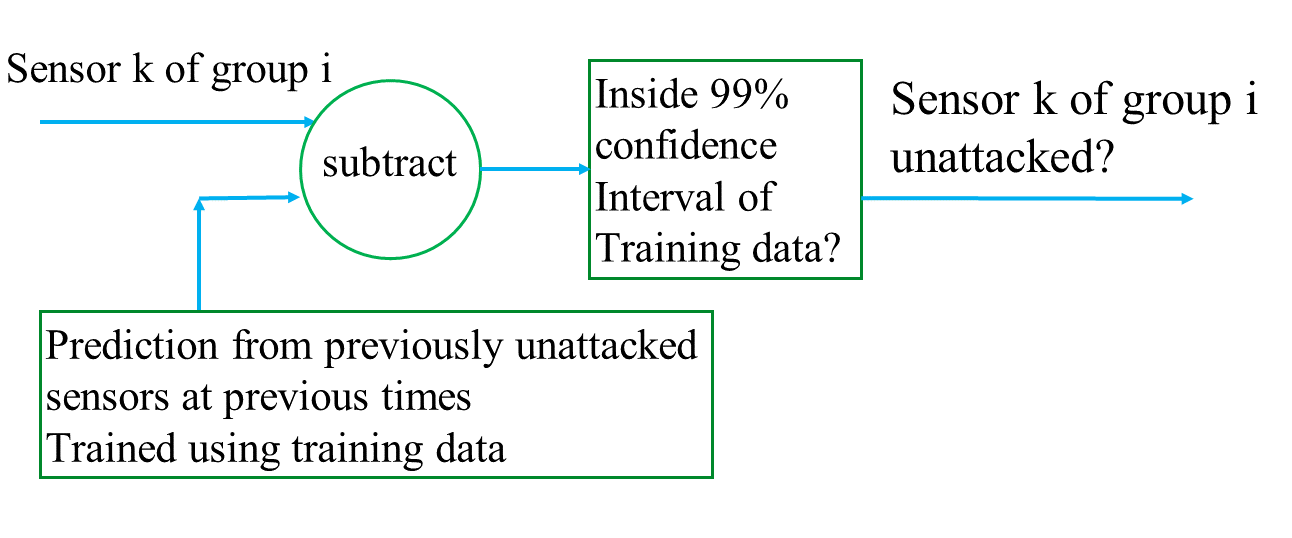 }
	\caption{{Block diagram} of APCC-SIMPLE with $\beta=99\%$.}
	\label{fAPCCS}
\end{figure}

The prediction of what a typical sensor output would look like at a given time step can employ any machine learning approach. 
In the numerical results, we focus on a prediction method based on the Random Forest approach as delineated by Breiman~\cite{breiman2001random}.
Since the number of sensors labeled as unattacked can vary over time, we employ multiple Random Forest models to accommodate different numbers of input features. Accordingly, when the number of unattacked sensors changes, a corresponding Random Forest model with the exact number of unattacked sensor inputs is selected to perform the prediction. For the Random Forest models, we use a 70/10/20 train/validate/test split.
Once the predictor is trained, 
the processing in APCC-SIMPLE is illustrated in Fig~\ref{fAPCCS} for $\beta=99\%$.

\subsection{APCC-ADDITIONAL}\label{APCCA}
For attackers that have knowledge of our protection approach, we augment the APCC-SIMPLE 
approach with additional processing in an approach we call APCC-ADDITIONAL.  Note that APCC-SIMPLE 
has already eliminated some sensors deemed to be attacked.  This additional processing 
may eliminate more.  To motivate the need for additional processing, consider the following. 
Suppose the malicious attackers know the details of the APCC-SIMPLE algorithm implemented to protect the sensor data. In that case, it's reasonable to infer that the attackers might insert many attacks (called edge attacks later) just inside one edge of the confidence interval shown in Fig~\ref{fAPCCS} to avoid the attacks from being identified while causing maximum damage. 
Thus the edge attacks are the worst-case attacks on APCC-SIMPLE. Since these attacks will not be identified without additional processing, they will skew the estimation of the trajectory.  If the attacker tries to launch a large number of such attacks, 
our additional processing  will 
attempt to eliminate many (or all) of these attacks by recognizing 
that they are not consistent with 
the statistics implied by the unattacked training data. Our approach will also mitigate other attacks. 

For additional protection, we employ a different method to ensure the deemed to be unattacked data is statistically close to the training data, assumed unattacked, when closeness is measured in a different manner.  The approach makes a decision on if a previously (based on APCC-SIMPLE) deemed unattacked sensor is attacked based on constructing the histogram of all the differences computed in Fig~\ref{fAPCCS}, at a given time and sensor group, that 
pass the APCC-SIMPLE test. 
We call this histogram
\(\hat{f}(x)\) when evaluated at a given bin $x$.  Note that a histogram is often called an 
empirical probability density function (PDF)  estimate and we use this to test if the statistics during operation match those obtained with training data in a specific sense. 

\begin{figure}[h!]
	\centering	\includegraphics[width=0.8\linewidth]{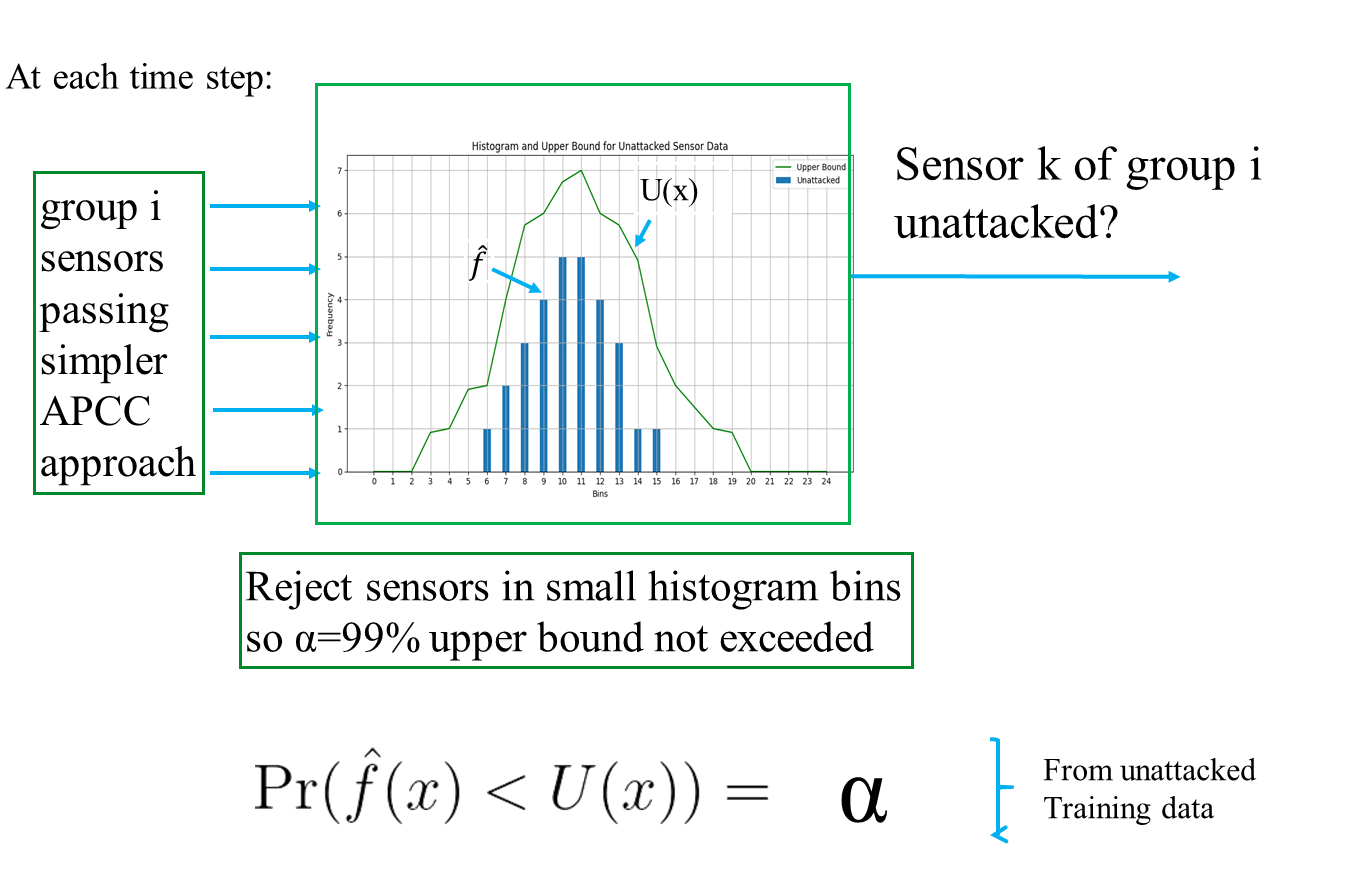 }
	\caption{Block diagram with $\alpha=99\%$ illustrating how the APCC-ADDITIONAL 
processing builds on 
the APCC-SIMPLE processing, resulting in a decision on if each sensor is unattacked at a given time.}
	\label{fAPCCA}
\end{figure}

In particular, APCC-ADDITIONAL uses the training data to find  the smallest upper limit of a  confidence interval, called $U(x)$, such that with probability $0 < \alpha < 1$ 
the histogram \(\hat{f}(x)\) should lie  below $U(x)$ 
as per \begin{equation}\label{confidenceinterval}
\Pr(\hat{f}(x) < U(x)) = \alpha.
\end{equation}
After learning $U(x)$ from the training data, 
we effectively eliminate sensor data which would cause \(\hat{f}(x)\) to exceed $U(x)$.  In particular, for any 
histogram bin where $\hat{f}(x) > U(x)$, we exclude the smallest number of sensors that produce values in that bin so afterwards $\hat{f}(x) \leq  U(x)$. 
The remaining sensor measurements, all those not excluded by either this or the previous processing, will be incorporated into the estimation/fusion processing as per Fig.~\ref{fAPCCA} and Fig.~\ref{f1}. 

\subsection{Worst-case Attacks}\label{WCA}

An important point is that we can calculate both the worst-case attack performance under the simple processing in APCC-SIMPLE (the edge attack mentioned previously in Subsection~\ref{APCCA}) and  
the worst-case attack performance under the additional processing in APCC-ADDITIONAL, a major contribution of our work. 
We will use this later in the numerical results to 
show the worst-case attack under APCC-ADDITIONAL can be made to cause significantly 
less damage that the 
worst-case attack under APCC-SIMPLE 
for cases with a large number of sensors being attacked, with the proper choice of the parameters in 
APCC-ADDITIONAL.   
If the worst-case attack can be tolerated, any attack can be tolerated. 

At any given time step, the optimum $N_A$-sensor attack against the 
APCC-ADDITIONAL approach 
will try to insert $N_A$  attacks whose magnitudes  lie in a set of histogram bins
which each provide 
a large enough difference 
between the upper bound $U(x)$ and the sensor data histogram $\hat{f}(x)$ to support those attacks (so the attacks get through) while choosing those bins that cause the greatest degradation to the estimation/fusion.   
Such attacks are called water-filling attacks since one can imagine pouring water to fill up the space left between the upper bound and the sensor data histogram, but the water must be poured in the most damaging bins first. 

\begin{figure}[h!]
	\centering	\includegraphics[width=0.8\linewidth]{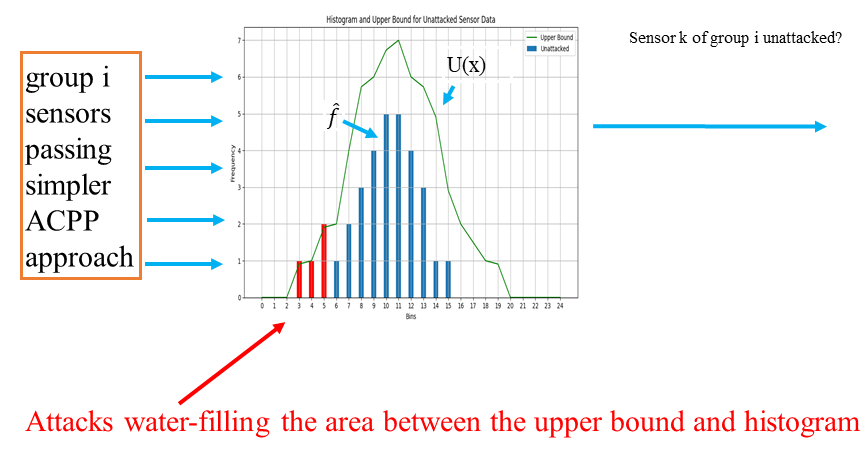 }
	\caption{{Block diagram} illustrating the worst-case water-filling attack for APCC-ADDITIONAL. The red lines show places where there is enough room between the upper bound and the sensor data histogram to insert one or two attacks and these attacks cause the greatest degradation to the estimation/fusion.   }
	\label{fWF}
\end{figure}

The approach is illustrated in Fig.~\ref{fWF}. 
The red lines show 
bins deemed to cause the most degradation to the estimation/fusion of all bins 
where there is enough room between the upper bound and the sensor data histogram to insert one or more attacks. 
As shown in Fig.~\ref{fWF}, one can only add attacks to a given histogram bin if there is 
enough space between the upper bound and the sensor data histogram.
This greatly limits the possible attacks which will pass this protection approach and the damage they can impose. 
Notice that increasing the number of sensors attacked beyond a certain value creates a smaller change in the 
possible degradation 
of the estimation/fusion for each additional sensor attacked  
since the additional sensors attacked  
have to eventually occupy bins which cause smaller degradation. 
At some point, an additional 
attacked sensor can no longer degrade the estimation/fusion  more 
than typical noise at that sensor. 

To find the particular $N_A$ sensors in which we will launch those attacks, we should pick those sensors that lead to the greatest damage, called {non-random} attacks.  Intuitively, these sensors will have unattacked sensor values that differ the most from the attack values.  Thus if the attack will be launched on the far left-hand side of the $x$ axis as in 
Fig.~\ref{fWF} (see red bins), then you should pick a sensor whose unattacked value is the farthest away, on the right-hand side of the $x$ axis in Fig.~\ref{fWF} (away from the red bins).

\section{Experimental Results}\label{sectionER} 

{
First, we remind the reader that, as explained at the end of Section~\ref{sec:LR}, 
EDAD 
approaches alone can yield very large, potentially  unbounded, errors for typical sequence estimation problems.  We provide numerical results of this type at the end of this section, where we also describe that by adding at least }{ APCC-SIMPLE} to the 
EDAD 
approaches, 
we can limit the errors to reasonably small values of our choosing. 

\begin{figure}[h!]
  \centering
  \includegraphics[width=\linewidth]{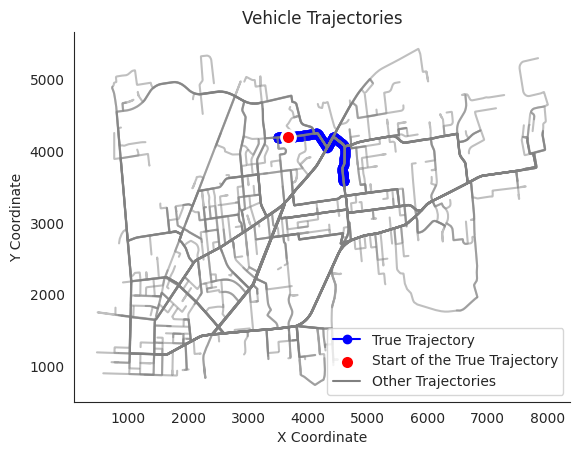}
  \caption{Map used with SUMO to produce trajectories.}
  \label{fig:map}
\end{figure}

{As CVN data sources with a sufficient number of sensors are unavailable and to test in a fully controllable environment, we generate sensor data used for experiments from a vehicle trajectory simulated using the Simulation of Urban Mobility open source software (SUMO) and introduce controllable sensor noise, independent samples from time to time. 
Most experiments will be
conducted for cases of both Gaussian and Laplacian sensor noise, two very different noise models \cite{poor}. Let the scale parameter 
of the Gaussian noise be $\sigma$ so that its variance is $\sigma^2$. Let the scale parameter 
of the Laplace noise be $b$ so that its variance is $2 b^2$ \cite{poor}. 
We vary the noise variance to characterize performance for different noise levels.}  Here we estimate a trajectory for a scalar quantity for simplicity, where the estimation/fusion algorithm uses only the sensor data at a given time step to produce the estimate at that same time step in the trajectory (no stored data) although we have considered other cases. 
We first employ SUMO to generate {192}  different possible noise-free vehicle path trajectories 
of length $m=150$ time steps 
that follow routes on a map shown in Fig~\ref{fig:map}.  
After adding independent noise from sensor to sensor  to these trajectories, we have our sensor data.  {In the examples, the sensor data describes  position.
Radar is an example of this case.   The noise added to each sensor has the same variance for easy interoperability.}
This sensor data will be used differently for testing APCC-SIMPLE and APCC-ADDITIONAL as discussed later. 

\begin{figure}[htbp]
  \centering
  \begin{subfigure}[b]{0.45\linewidth}
    \centering
    \includegraphics[width=\linewidth]{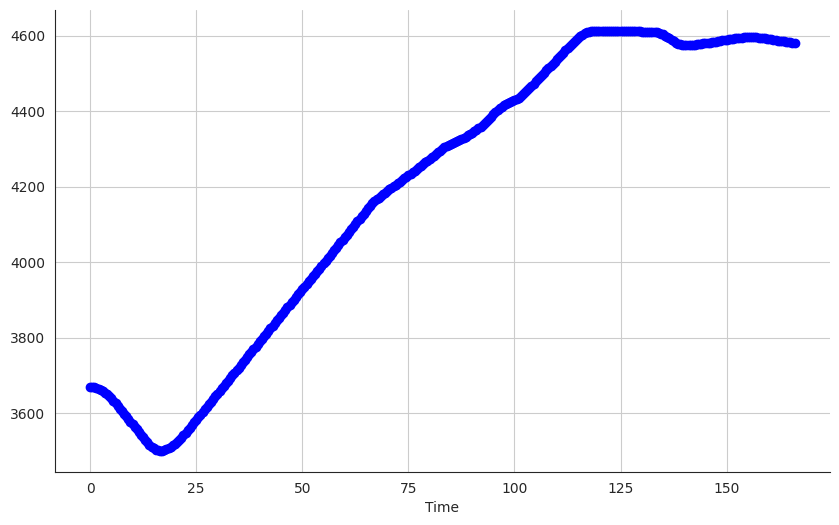}
    \caption{$x$ coordinate waveform of the true trajectory.}
    \label{fig:p1}
  \end{subfigure}
  \hfill
  \begin{subfigure}[b]{0.45\linewidth}
    \centering
    \includegraphics[width=\linewidth]{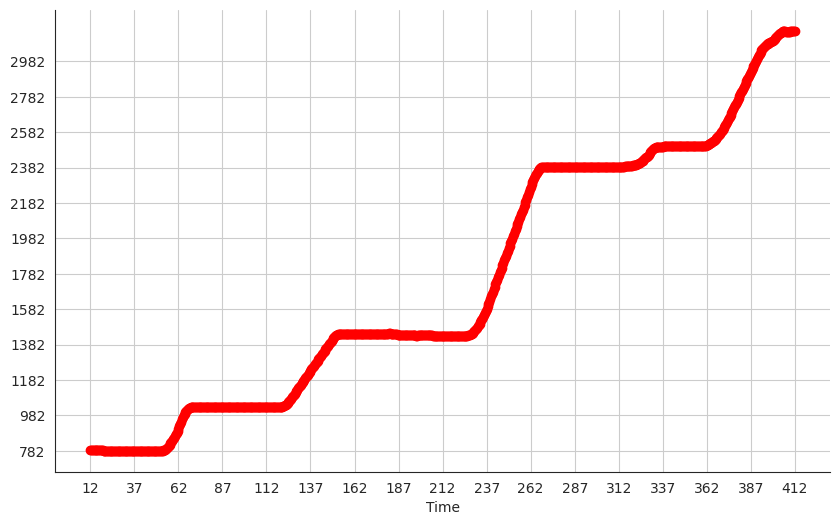}
    \caption{$x$ coordinate waveform of an attacked trajectory.}
    \label{fig:p2}
  \end{subfigure}
  \caption{Comparison of true and attacked trajectories.}
  \label{fig:waveforms}
\end{figure}


While our protection approach allows changing the number of sensors attacked each time step, here we fix the number of sensors attacked each time step for ease of interpretation of the results and call the number of attacked sensors $N_{A}$.  
Here $N_A < N$ where $N$ is the total number of sensors available (attacked plus unattacked).
In order to focus on the ability of our approach to find and eliminate attacked sensor data and to remove any effects due to the estimation/fusion algorithm in our numerical results, we employ Maximum Likelihood Estimation (MLE)~\cite{poor,Kay} as the fusion approach, which is optimized under the assumption of known sensor noise distribution.  
For Gaussian noise or a large number of provided sensor observations, the MLE is optimum among unbiased
estimators. 
We have observed that these MLE results can well approximate performance for highly optimized machine learning-based fusion algorithms.  We have also found that non-optimized 
machine learning-based 
 fusion algorithms do impose some additional degradation in performance, while we do not provide these results here.  

None of the  
attacks we consider in our tests 
would be detected by pure EDAD since these attacks substitute one valid 
sequence for another as discussed in the  second to last paragraph of
Section~\ref{sec:LR}.  Our purpose here is to show our new approaches can achieve very good performance for such attacks, thus providing a useful way of augmenting EDAD to perform well for attacks that EDAD  misses. 
We should note that for the kind of attacks we consider, it’s obvious that using attacked data (rather than eliminating it) will never improve performance since these attacks delete all the useful information about the original trajectory
by changing it. 

{ 
Let the actual trajectory we wish to estimate be 
$ y_1,\ldots,y_m $. Let the vector of 
sensor observations at time step $j, j=1,\ldots, m$ be $ X_j = ( x_{1j}, \ldots, x_{Nj})^T $. The estimation/fusion algorithm will use this vector to produce 
an estimate for time step $j$ which we denote as 
$ \hat{y}_j $.  
For any given estimation/fusion method, the performance is measured using 
the mean square error (MSE), 
defined as 

\begin{equation}
\text{MSE} = \mathbb{E} \left[\frac{1}{m}\sum_{j=1}^{m} ( \hat{y}_j 
 - y_j )^2\right].
 \label{MSE}
\end{equation} 
The Normalized Root MSE~(NRMSE) can be defined as
\begin{equation}
\text{NRMSE} = \sqrt{\frac{\text{MSE}}{\frac{1}{m}\sum_{j=1}^{m}|y_j|}} = \sqrt{\frac{\mathbb{E} \left[\frac{1}{m}\sum_{j=1}^{m} ( \hat{y}_j - y_j )^2\right]}{\frac{1}{m}\sum_{j=1}^{m}|y_j|}}
\label{NRMSE}
\end{equation}
}
By employing an optimized estimate which also only uses  the sensor data which is actually unattacked, we obtain a useful performance comparison that incorporates information not normally available in practice called the   
{Genie Estimate} (GE), The GE sets $\hat{y}_j$ in (\ref{MSE}) as the MLE estimate/fusion for the known noise distribution using 
only the unattacked sensor data in $ X_j$.  

{
Here we describe more details on the construction procedure for $U(x)$ that we used in the numerical results.  We use  unattacked sensor training data, generated from SUMO with noise added. 
As in all machine learning  processing, more training data is better to give accurate estimates of $U(x)$. In our experiments we found 
the approach does seem to work well with as little as 37500 labeled training data samples describing the sensor observations.  This training data is processed as per Fig. 4 and Fig. 5, giving a set of data which describes many realizations of the after-histogram data.  Using it, we can calculate 
the smallest value in each bin $x$, call it $U(x)$, such that approximately $\alpha$ percent of the  total amount of data 
that is in the bin is below $U(x)$ for any desired $\alpha$. }

{
We have found, through experiments, that bins of width  $21\%$ as a percentage of the standard deviation of the sensor output work well.  Slightly smaller bins also work well.  
}

\subsection{Numerical Results for APCC-SIMPLE: Attacks Without Knowledge}\label{APCCSIMPLE}

In this subsection, we discuss the numerical results obtained with  the APCC-SIMPLE algorithm, described in Subsection~\ref{APCCS}. In the experiments, the chosen $\beta$ is 99.9\%, the total number of time steps in the trajectory $m$ is 150, and the sampling period of the sensor data is $1.0 \times 10^{-3}$ sec. To obtain an accurate estimate of  NRMSE in (\ref{NRMSE}), we performed a Monte Carlo (MC) simulation where we average the NRMSE over 
1 million independent 
realizations called the MC run length.

We consider cases where 
the attacker is assumed to not have detailed knowledge of the protection scheme, for example the confidence interval in Fig.~\ref{fAPCCS} or the prediction. Thus the attacker is unable to launch optimum (worst case) attacks that are all focused to lie at the edge of the confidence interval in Fig.~\ref{fAPCCS}.  We consider such cases later. 
Instead, for ease in comparing the various cases considered in this subsection, the noise-free unattacked trajectory was chosen to be the 
trajectory shown in  Figure~\ref{fig:p1}, which is 
one of the trajectories  from the set of those 
generated using the map 
in Fig~\ref{fig:map}. 
The noise-free attacked trajectory 
was chosen randomly (attacker can not optimize) from the other waveforms generated using the map 
in Fig~\ref{fig:map}.
One of these possible attacked trajectories is shown in  Figure~\ref{fig:p2}.  The attacked trajectories are chosen randomly and independently at each attacked sensor without replacement.  


In Table~\ref{simpleG} and Table~\ref{SimpleL}, we show the numerical results for various cases we have tested. The numerical results illustrate that for these cases where we apply the APCC-SIMPLE approach with optimized MLE fusion to address 
attackers without protection system knowledge, the NRMSE between the APCC-SIMPLE (SIMPLE in Tables) 
 estimated trajectory and the true trajectory 
(last column) is small and it is very close (identical to two decimal places) 
to  that for the GE (using MLE). It indicates the APCC-SIMPLE approach successfully identifies the attacked sensors and excludes their readings accordingly in the estimation/fusion process. 


\begin{table}[htb]
\centering
\caption{APCC-SIMPLE Gaussian Results $\beta = 99.9\%$ 
{ \small \\$\sigma^2$ is noise variance, $N$ is $\#$ of sensors, $N_A$ is  $\#$ of attacked sensors,  NRMSE GE and NRMSE SIMPLE (APCC-SIMPLE)  both use MLE. $m=150.$}}
\label{simpleG}
\begin{tabular}{|c|c|c|c|c|}
\hline
\(\sigma^2\) & \(N\) & \(N_{A}\) & \(\text{NRMSE}\) GE & \(\text{NRMSE}\) SIMPLE \\
\hline
\(1.0 \times 10^{-4}\) & 50 & 40 & $5.03 \times 10^{-5}$ & $5.03 \times 10^{-5}$ \\
\(1.0 \times 10^{-4}\) & 50 & 30 & $3.56 \times 10^{-5}$ & $3.56 \times 10^{-5}$ \\
\(1.0 \times 10^{-4}\) & 50 & 10 & $2.52 \times 10^{-5}$ & $2.52 \times 10^{-5}$ \\
\(1.0 \times 10^{-4}\) & 20 & 10 & $5.03 \times 10^{-5}$ & $5.03 \times 10^{-5}$ \\
\(1.0 \times 10^{-4}\) & 10 & 5 & $7.12 \times 10^{-5}$ & $7.12 \times 10^{-5}$ \\
\(1.0 \times 10^{-2}\) & 50 & 40 & $5.03 \times 10^{-4}$ & $5.03 \times 10^{-4}$ \\
\(1.0 \times 10^{-2}\) & 50 & 30 & $3.56 \times 10^{-4}$ & $3.56 \times 10^{-4}$ \\
\(1.0 \times 10^{-2}\) & 50 & 10 & $2.52 \times 10^{-4}$ & $2.52 \times 10^{-4}$ \\
\(1.0 \times 10^{-2}\) & 20 & 10 & $5.03 \times 10^{-4}$ & $5.03 \times 10^{-4}$ \\
\(1.0 \times 10^{-2}\) & 10 & 5 & $7.12 \times 10^{-4}$ & $7.12 \times 10^{-4}$ \\
\hline
\end{tabular}
\end{table}

\begin{table}[htb]
\centering
\caption{APCC-SIMPLE Laplacian Results $\beta = 99.9\%$ \\
{\small See TABLE~\ref{simpleG}
 for column definitions. $m=150.$}} 
 \label{SimpleL}
\begin{tabular}{|c|c|c|c|c|}
\hline
\( \sigma^2=2b^2\) & \(N\) & \(N_{A}\) & \(\text{NRMSE}\) GE & \(\text{NRMSE}\) SIMPLE \\
\hline
\(2.0 \times 10^{-4}\) & 50 & 40 & $6.06 \times 10^{-5}$ & $6.06 \times 10^{-5}$ \\
\(2.0 \times 10^{-4}\) & 50 & 30 & $4.11 \times 10^{-5}$ & $4.11 \times 10^{-5}$ \\
\(2.0 \times 10^{-4}\) & 50 & 10 & $2.80 \times 10^{-5}$ & $2.80 \times 10^{-5}$ \\
\(2.0 \times 10^{-4}\) & 20 & 10 & $6.06 \times 10^{-5}$ & $6.06 \times 10^{-5}$ \\
\(2.0 \times 10^{-4}\) & 10 & 5  & $9.43 \times 10^{-5}$ & $9.43 \times 10^{-5}$ \\
\(2.0 \times 10^{-2}\) & 50 & 40 & $6.06 \times 10^{-4}$ & $6.06 \times 10^{-4}$ \\
\(2.0 \times 10^{-2}\) & 50 & 30 & $4.11 \times 10^{-4}$ & $4.11 \times 10^{-4}$ \\
\(2.0 \times 10^{-2}\) & 50 & 10 & $2.80 \times 10^{-4}$ & $2.80 \times 10^{-4}$ \\
\(2.0 \times 10^{-2}\) & 20 & 10 & $6.06 \times 10^{-4}$ & $6.06 \times 10^{-4}$ \\
\(2.0 \times 10^{-2}\) & 10 & 5 & $9.43 \times 10^{-4}$ & $9.43 \times 10^{-4}$ \\
\hline
\end{tabular}
\end{table}

\begin{table}[htb]
\centering
\caption{
=
APCC-SIMPLE Gaussian with $m=4.8 \times 10^{5}$. \\
See TABLE I for column definitions.
}

\label{simpleGlong}

\begin{tabular}{|c|c|c|c|c|}
\hline
$\sigma^2$ & $N$ & $N_{A}$ & $\text{NRMSE GE}$ & $\text{NRMSE SIMPLE}$ \\
\hline
$1.0 \times 10^{-4}$ & 50 & 40 & $5.19 \times 10^{-5}$ & $5.19 \times 10^{-5}$ \\
$1.0 \times 10^{-4}$ & 50 & 30 & $3.67 \times 10^{-5}$ & $3.67 \times 10^{-5}$ \\
$1.0 \times 10^{-4}$ & 50 & 10 & $2.60 \times 10^{-5}$ & $2.60 \times 10^{-5}$ \\
$1.0 \times 10^{-4}$ & 20 & 10 & $5.19 \times 10^{-5}$ & $5.19 \times 10^{-5}$ \\
$1.0 \times 10^{-4}$ & 10 & 5  & $7.34 \times 10^{-5}$ & $7.34 \times 10^{-5}$ \\
$1.0 \times 10^{-2}$  & 50 & 40 & $5.19 \times 10^{-4}$  & $5.19 \times 10^{-4}$ \\
$1.0 \times 10^{-2}$  & 50 & 30 & $3.67 \times 10^{-4}$ & $3.67 \times 10^{-4}$ \\
$1.0 \times 10^{-2}$  & 50 & 10 & $2.60 \times 10^{-4}$  & $2.60 \times 10^{-4}$ \\
$1.0 \times 10^{-2}$  & 20 & 10 & $5.19 \times 10^{-4}$ & $5.19 \times 10^{-4}$ \\
$1.0 \times 10^{-2}$  & 10 & 5  & $7.34 \times 10^{-4}$ & $7.34 \times 10^{-4}$ \\
\hline
\end{tabular}

\end{table}

\begin{table}[htb]
\centering
\caption{{APCC-SIMPLE Laplacian with $m=4.8 \times 10^{5}$. \\ See TABLE I for column definitions.}} 
\label{LaplaceL}

\begin{tabular}{|c|c|c|c|c|}
\hline
\( \sigma^2=2b^2\) & \(N\) & \(N_{A}\) & \(\text{NRMSE}\) GE & \(\text{NRMSE}\) SIMPLE \\
\hline
\(2.0 \times 10^{-4}\) & 50 & 40 & $6.25 \times 10^{-5}$ & $6.25 \times 10^{-5}$ \\
\(2.0 \times 10^{-4}\) & 50 & 30 & $4.24 \times 10^{-5}$ & $4.24 \times 10^{-5}$ \\
\(2.0 \times 10^{-4}\) & 50 & 10 & $2.89 \times 10^{-5}$ & $2.89 \times 10^{-5}$ \\
\(2.0 \times 10^{-4}\) & 20 & 10 & $6.25 \times 10^{-5}$ & $6.25 \times 10^{-5}$ \\
\(2.0 \times 10^{-4}\) & 10 & 5  & $9.73 \times 10^{-5}$ & $9.73 \times 10^{-5}$ \\
\(2.0 \times 10^{-2}\) & 50 & 40 & $6.25 \times 10^{-4}$ & $6.25 \times 10^{-4}$ \\
\(2.0 \times 10^{-2}\) & 50 & 30 & $4.24 \times 10^{-4}$ & $4.24 \times 10^{-4}$ \\
\(2.0 \times 10^{-2}\) & 50 & 10 & $2.89 \times 10^{-4}$ & $2.89 \times 10^{-4}$ \\
\(2.0 \times 10^{-2}\) & 20 & 10 & $6.25 \times 10^{-4}$ & $6.25 \times 10^{-4}$ \\
\(2.0 \times 10^{-2}\) & 10 & 5 & $9.73 \times 10^{-4}$ & $9.73 \times 10^{-4}$ \\
\hline
\end{tabular}
\end{table}

In Table~\ref{simpleGlong}, we performed exactly the same experiment as in Table~\ref{simpleG}, but used a true trajectory (similar to Figure~\ref{fig:p1}) that is longer, with $m=4.8 \times 10^{5}$ instead of $m=150$. Since the true trajectory is different, as expected from (\ref{NRMSE}), the normalization factor in the NRMSE changed. This change makes the GE slightly larger than in Table~\ref{simpleG}, but the change is just as expected given the new true trajectory. 
According to the results shown in Table~\ref{simpleGlong}, our APCC-SIMPLE approach yields NRMSE values identical to those of GE with two decimal places in all tested cases. For example, when 40 out of 50 sensors are under attack and the variance of the Gaussian noise is $1.0 \times 10^{-4}$, the NRMSE of the APCC-SIMPLE approach is $5.19 \times 10^{-5}$, which matches the NRMSE of the GE in this new case to two decimal places, which proves our APCC-SIMPLE approach works well in this case.  

Similar results for Laplacian noise are shown in Table~\ref{LaplaceL}, which performs exactly the same experiment as conducted in Table~\ref{simpleGlong} but changes the noise distribution from Gaussian to Laplacian. According to the results shown in Table~\ref{LaplaceL}, our APCC-SIMPLE approach produces NRMSE values identical to those of GE to  two decimal places in all tested cases. For instance, when 40 out of 50 sensors are under attack and the variance of the Laplacian noise is $2.0 \times 10^{-4}$, the NRMSE of the APCC-SIMPLE approach is $6.25 \times 10^{-5}$, which matches the NRMSE of the GE in this new case to two decimal places. Thus our APCC-SIMPLE approach works well in this case also.  

In the next few results to follow, we go back to using $m=150$ unless otherwise stated. We defined $m$ as the trajectory length. Why don't you use this? Making $m$ longer is easy to explain. Refer back to (2). In Table~\ref{simpleGlong} we conduct the same experiment as in Table~\ref{simpleG}, but using a true waveform of much longer duration while keeping otherwise identical settings. The numerical results in Table~\ref{simpleGlong} confirm that our APCC-SIMPLE approach performs reliably under this extended scenario. In particular, the NRMSE of GE remains consistently close to that of APCC-SIMPLE, demonstrating that the proposed method maintains robust performance even for long waveforms.

\subsection{Numerical Results for APCC-ADDITIONAL and Attacks With Knowledge }\label{APCCADDITIONAL}

\subsubsection{Edge Attack -  APCC-SIMPLE vs APCC-ADDITIONAL }

In this subsection, we consider 
the worst-case attack, called an edge attack, for the 
APCC-SIMPLE approach and demonstrate the performance improvements obtainable for such attacks when the APCC-ADDITIONAL approach is applied. 
The edge attack places the 
attacked sensor values just  inside the edge of the confidence interval shown in Fig~\ref{fAPCCS}, thus they require knowledge of the protection approach to launch. In the experiments, the number of sensors $N$ is fixed to be 50 while we vary the number of attacked sensors $N_A$ to investigate the impact of attacking more sensors. The total number of time steps considered $m$ in the trajectory to be estimated is still 150. To perform the APCC-ADDITIONAL processing, we use histograms of 25 bins to calculate $U(x)$ and \(\hat{f}(x)\) in Fig.~\ref{fAPCCA}.  We consider two specific types 
of edge attacks.  One where we select the sensors to attack randomly, called a random attack, and the other where we select the sensors to attack which will cause the most damage, called a nonrandom attack.  
Since we pick the attacks near one edge of the confidence interval 
in Fig~\ref{fAPCCS}, the  sensors which will cause the most damage are those with unattacked data closest to the other end of the confidence interval while lying inside the confidence interval. 

\begin{table*}[htb]
\centering
\caption{Gaussian Non-Random Edge Attack $\alpha = 90\%$ $\beta = 99.9\%$ $m=150.$}
\label{ApccAEdgeNotrandomAttack}
\begin{tabular}{|c|c|c|c|c|c|c|}
\hline
Variance & approach &  \(N_{A} = 0\) & \(N_{A} = 10\) & \(N_{A} = 20\) & \(N_{A} = 30\) & \(N_{A} = 40\) \\
\hline
\multirow{3}{*}{\(1.0 \times 10^{-4}\)} & NRMSE GE & $2.25 \times 10^{-5}$ & \(5.93 \times 10^{-5}\) &  \(1.04 \times 10^{-4}\) & \(1.54 \times 10^{-4}\) & \(2.21 \times 10^{-4}\) \\
\cline{2-7}
 & NRMSE APCC-SIMPLE & $2.25 \times 10^{-5}$ & \(2.10 \times 10^{-4}\) & \(4.09 \times 10^{-4}\) & \(6.12 \times 10^{-4}\) & \(8.34 \times 10^{-4}\) \\
\cline{2-7}
 & NRMSE APCC-ADDITIONAL & $2.28 \times 10^{-5}$ & \(6.65 \times 10^{-5}\) & \(1.09 \times 10^{-4}\) &  \(1.56 \times 10^{-4}\) & \(2.21 \times 10^{-4}\) \\
\hline
\multirow{3}{*}{\(1.0 \times 10^{-2}\)} & NRMSE GE & $2.25 \times 10^{-4}$ & \(5.93 \times 10^{-4}\) & \(1.04 \times 10^{-3}\) & \(1.54 \times 10^{-3}\) & \(2.21 \times 10^{-3}\) \\
\cline{2-7}
 & NRMSE APCC-SIMPLE & $2.25 \times 10^{-4}$ & \(2.06 \times 10^{-3}\) & \(4.02 \times 10^{-3}\) & \(6.02 \times 10^{-3}\) & \(8.19 \times 10^{-3}\) \\
\cline{2-7}
 & NRMSE APCC-ADDITIONAL & \(2.28 \times 10^{-4}\) & \(6.08 \times 10^{-4}\) & \(1.05 \times 10^{-3}\) & \(1.54 \times 10^{-3}\) & \(2.21 \times 10^{-3}\) \\
\hline
\end{tabular}
\vspace{0.1cm}

\textit{Note: The NRMSEs in this table are obtained by using MLE as the fusion approach.}
\end{table*}

\begin{table*}[htb]
\centering
\caption{Gaussian Random Edge Attack $\alpha = 90\%$ $\beta = 99.9\%$ $m=150.$}
\label{ApccAEdgerandomAttackGaussian}
\begin{tabular}{|c|c|c|c|c|c|c|}
\hline
Variance & approach & \(N_{A} = 0\) & \(N_{A} = 10\) & \(N_{A} = 20\) &  \(N_{A} = 30\) & \(N_{A} = 40\) \\
\hline
\multirow{3}{*}{\(1.0 \times 10^{-4}\)} & NRMSE GE & $2.25 \times 10^{-5}$ &   $2.52 \times 10^{-5}$ & $2.91 \times 10^{-5}$ &  $3.56 \times 10^{-5}$ & $5.03 \times 10^{-5}$  \\
\cline{2-7}
 & NRMSE APCC-SIMPLE & $2.25 \times 10^{-5}$ & $1.44 \times 10^{-4}$ & $2.86 \times 10^{-4}$ & $4.29 \times 10^{-4}$ & $5.69 \times 10^{-4}$  \\
\cline{2-7}
 & NRMSE APCC-ADDITIONAL & $2.28 \times 10^{-5}$ &$2.52 \times 10^{-5}$ & $2.91 \times 10^{-5}$ & $3.56 \times 10^{-5}$ & $5.03 \times 10^{-5}$  \\
\hline
\multirow{3}{*}{\(1.0 \times 10^{-2}\)} & NRMSE GE & $2.25 \times 10^{-4}$  & $2.52 \times 10^{-4}$ & $2.91 \times 10^{-4}$ & $3.56 \times 10^{-4}$ & $5.03 \times 10^{-4}$  \\
\cline{2-7}
 & NRMSE APCC-SIMPLE & $2.25 \times 10^{-4}$  & $1.43 \times 10^{-3}$  &   $2.87 \times 10^{-3}$ & $4.28 \times 10^{-3}$  & $5.73 \times 10^{-3}$ \\
\cline{2-7}
 & NRMSE APCC-ADDITIONAL & $2.28 \times 10^{-4}$ & $2.52 \times 10^{-4}$  & $2.91 \times 10^{-4}$ & $3.56 \times 10^{-4}$  &  $5.03 \times 10^{-4}$\\
\hline
\end{tabular}
\vspace{0.1cm}

\textit{Note: The NRMSEs in this table are obtained by using MLE as the fusion approach.}
\end{table*}

For the case when the sensor noise is Gaussian, 
Table~\ref{ApccAEdgeNotrandomAttack} (Non-Random attack) and Table~\ref{ApccAEdgerandomAttackGaussian} (Random attack) show the 
NRMSE performance improvements of the APCC-ADDITIONAL approach over the APCC-SIMPLE approach for the edge attacks.  
For large $N_A$, the APCC-ADDITIONAL approach nearly perfectly identifies and removes all the edge attacked sensors to provide NRMSE performance close to the GE 
(identical to two decimal places).  
The performance of the APCC-ADDITIONAL approach is considerably better than that of the 
performance of the APCC-SIMPLE approach for the edge attacks which is extremely important in some critical applications.  However, considering its simplicity. the performance of the APCC-SIMPLE approach for the edge attacks is not that bad 
and might be suitable in some noncritical applications. 
Note that the degradation of the worst-case attack (Edge attack) is bounded to a reasonably small value for APCC-SIMPLE but this will not be the case if we remove the APCC-SIMPLE processing as discussed previously. 
In the cases considered,   the nonrandom attacks are more powerful than the corresponding random attacks as expected. 
Similar conclusions to those drawn from 
Table~\ref{ApccAEdgeNotrandomAttack} and Table~\ref{ApccAEdgerandomAttackGaussian}  
can be drawn from the cases shown in Table~\ref{ApccAEdgenonrandomAttackLaplace} (NonRandom attack) and Table~\ref{ApccAEdgerandomAttackLaplace} (Random attack) which consider cases with Laplacian noise as opposed to Gaussian noise.

\begin{table*}[htb]
\centering
\caption{Laplacian Non-Random Edge Attack $\alpha = 90\%$ $\beta = 99.9\%$ $m=150.$}
\label{ApccAEdgenonrandomAttackLaplace}
\begin{tabular}{|c|c|c|c|c|c|c|}
\hline
Variance & approach & \(N_{A} = 0\) & \(N_{A} = 10\) & \(N_{A} = 20\) & \(N_{A} = 30\) & \(N_{A} = 40\) \\
\hline
\multirow{3}{*}{\(2.0 \times 10^{-4}\)} & NRMSE GE & $2.48 \times 10^{-5}$ & $4.62 \times 10^{-5}$ & $8.97 \times 10^{-5}$ & $1.54 \times 10^{-4}$ & $2.66 \times 10^{-4}$  \\
\cline{2-7}
 & NRMSE APCC-SIMPLE & $2.48 \times 10^{-5}$ & $8.88 \times 10^{-5}$  & $2.65 \times 10^{-4}$ & $1.55 \times 10^{-3}$ & $1.66 \times 10^{-3}$ \\
\cline{2-7}
 & NRMSE APCC-ADDITIONAL & $2.48 \times 10^{-5}$ & $4.85 \times 10^{-5}$ & $9.11 \times 10^{-5}$  &  $1.54 \times 10^{-4}$ &  $2.66 \times 10^{-4}$  \\
\hline
\multirow{3}{*}{\(2.0 \times 10^{-2}\)} & NRMSE GE & $2.48 \times 10^{-4}$  & $4.62 \times 10^{-4}$ & $8.97 \times 10^{-4}$ & $1.54 \times 10^{-3}$ &  $2.66 \times 10^{-3}$ \\
\cline{2-7}
 & NRMSE APCC-SIMPLE & $2.48 \times 10^{-4}$  & $8.90 \times 10^{-4}$ &  $2.65 \times 10^{-3}$ &  $1.52 \times 10^{-2}$ &  $1.63 \times 10^{-2}$ \\
\cline{2-7}
 & NRMSE APCC-ADDITIONAL & $2.48 \times 10^{-4}$ & $4.78 \times 10^{-4}$  & $9.03 \times 10^{-4}$  &  $1.54 \times 10^{-3}$ &  $2.66 \times 10^{-3}$  \\
\hline
\end{tabular}
\vspace{0.1cm}

\textit{Note: The NRMSEs in this table are obtained by using MLE as the fusion approach.}
\end{table*}

\begin{table*}[htb]
\centering
\caption{Laplacian Random Edge Attack $\alpha = 90\%$ $\beta = 99.9\%$ $m=150.$}
\label{ApccAEdgerandomAttackLaplace}
\begin{tabular}{|c|c|c|c|c|c|c|}
\hline
Variance & approach & \(N_{A} = 0\) & \(N_{A} = 10\) & \(N_{A} = 20\) & \(N_{A} = 30\) & \(N_{A} = 40\) \\
\hline
\multirow{3}{*}{\(2.0 \times 10^{-4}\)} & NRMSE GE & $2.48 \times 10^{-5}$ &  $2.80 \times 10^{-5}$ & $3.29 \times 10^{-5}$ & $4.11 \times 10^{-5}$  & $6.06 \times 10^{-5}$  \\
\cline{2-7}
 & NRMSE APCC-SIMPLE & $2.48 \times 10^{-5}$ & $5.76 \times 10^{-5}$ & $1.88 \times 10^{-4}$ &  $1.35 \times 10^{-3}$ &   $1.35 \times 10^{-3}$ \\
\cline{2-7}
 & NRMSE APCC-ADDITIONAL & $2.48 \times 10^{-5}$ & $2.80 \times 10^{-5}$ & $3.29 \times 10^{-5}$  & $4.11 \times 10^{-5}$  &  $6.06 \times 10^{-5}$  \\
\hline
\multirow{3}{*}{\(2.0 \times 10^{-2}\)} & NRMSE GE & $2.48 \times 10^{-4}$  & $2.80 \times 10^{-4}$ & $3.29 \times 10^{-4}$ & $4.11 \times 10^{-4}$ & $6.06 \times 10^{-4}$   \\
\cline{2-7}
 & NRMSE APCC-SIMPLE & $2.48 \times 10^{-4}$  & $5.73 \times 10^{-4}$ &  $1.88 \times 10^{-3}$  & $1.36 \times 10^{-2}$  &   $1.37 \times 10^{-2}$ \\
\cline{2-7}
 & NRMSE APCC-ADDITIONAL & $2.48 \times 10^{-4}$  & $2.80 \times 10^{-4}$ &$3.29 \times 10^{-4}$ & $4.11 \times 10^{-4}$  & $6.06 \times 10^{-4}$  \\
\hline
\end{tabular}
\vspace{0.1cm}

\textit{Note: The NRMSEs in this table are obtained by using MLE as the fusion approach.}
\end{table*}

\begin{table*}[htb]
\centering
\caption{Gaussian Optimized Non-Random Water Filling Attack. $m=150.$}
\label{ApccAWFNonrandomGasussian}
\begin{tabular}{|c|c|c|c|c|c|c|}
\hline
Variance & approach & \(N_{A} = 0\) & \(N_{A} = 10\) & \(N_{A} = 20\) & \(N_{A} = 30\) & \(N_{A} = 40\) \\
\hline
\multirow{6}{*}{\(1.0 \times 10^{-4}\)} 
 & Worst-case for SIMPLE ($\beta = 99.9\%$) & \textendash
 & \(2.10 \times 10^{-4}\) & \(4.09 \times 10^{-4}\) & \(6.12 \times 10^{-4}\) & \(8.34 \times 10^{-4}\) \\
\cline{2-7}
 & NRMSE for $\alpha = 80\%$, $\beta = 80\%$ & \(4.15 \times 10^{-5}\) & \(1.29 \times 10^{-4}\) & \(2.56 \times 10^{-4}\) & \(4.03 \times 10^{-4}\) & \(4.47 \times 10^{-4}\) \\
\cline{2-7}
 & NRMSE for $\alpha = 70\%$, $\beta = 80\%$ & \(4.94 \times 10^{-5}\) & \(1.29 \times 10^{-4}\) &  \(2.19 \times 10^{-4}\)& \(2.60 \times 10^{-4}\) &  \(2.60 \times 10^{-4}\) \\
\cline{2-7}
 & NRMSE for $\alpha = 60\%$, $\beta = 80\%$ & \(5.24 \times 10^{-5}\) & \(1.29 \times 10^{-4}\) & \(1.86 \times 10^{-4}\) & \(1.86 \times 10^{-4}\) & \(1.86 \times 10^{-4}\) \\
\cline{2-7}
 & NRMSE for $\alpha = 50\%$, $\beta = 80\%$ & \(6.06 \times 10^{-5}\) & \(1.22 \times 10^{-4}\) & \(1.40 \times 10^{-4}\) & \(1.40 \times 10^{-4}\) & \(1.40 \times 10^{-4}\) \\
\cline{2-7}
 & NRMSE for $\alpha = 40\%$, $\beta = 80\%$ & \(7.14 \times 10^{-5}\) & \(9.31 \times 10^{-5}\)  & \(9.33 \times 10^{-5}\) & \(9.33 \times 10^{-5}\) & \(9.33 \times 10^{-5}\) \\
\hline
\multirow{6}{*}{\(1.0 \times 10^{-2}\)} 
 & Worst-case for SIMPLE (for $\beta = 99.9\%$) & \textendash
 & \(2.06 \times 10^{-3}\) & \(4.02 \times 10^{-3}\) & \(6.02 \times 10^{-3}\) & \(8.19 \times 10^{-3}\) \\
\cline{2-7}
 & NRMSE for $\alpha = 80\%$, $\beta = 80\%$ & \(2.45 \times 10^{-4}\) &   \(1.29 \times 10^{-3}\) & \(2.74 \times 10^{-3}\) & \(4.79 \times 10^{-3}\) & \(5.53 \times 10^{-3}\)  \\
\cline{2-7}
 & NRMSE for $\alpha = 70\%$, $\beta = 80\%$ & \(2.70 \times 10^{-4}\) & \(1.29 \times 10^{-3}\) & \(2.52 \times 10^{-3}\) & \(3.12 \times 10^{-3}\) & \(3.12 \times 10^{-3}\) \\
\cline{2-7}
 & NRMSE for $\alpha = 60\%$, $\beta = 80\%$ & \(2.91 \times 10^{-4}\) & \(1.29 \times 10^{-3}\)  & \(1.82 \times 10^{-3}\)  & \(1.82 \times 10^{-3}\)  & \(1.82 \times 10^{-3}\) \\
\cline{2-7}
 & NRMSE for $\alpha = 50\%$, $\beta = 80\%$ &  \(3.34 \times 10^{-4}\) & \(9.17 \times 10^{-4}\) & \(9.54 \times 10^{-4}\) &  \(9.48 \times 10^{-4}\) &  \(9.48 \times 10^{-4}\) \\
\cline{2-7}
 & NRMSE for $\alpha = 40\%$, $\beta = 80\%$ &  \(3.91 \times 10^{-4}\) & \(5.60 \times 10^{-4}\) & \(5.60 \times 10^{-4}\) & \(5.60 \times 10^{-4}\) &  \(5.60 \times 10^{-4}\) \\
\hline
\end{tabular}
\vspace{0.1cm}

\textit{Note: The NRMSEs in this table are obtained by using MLE as the fusion approach.}
\end{table*}

\begin{table*}[htb]
\centering
\caption{Laplacian Optimized Non-Random Water Filling Attack. $m=150.$}
\label{ApccAWFNonrandomLaplacian60}
\begin{tabular}{|c|c|c|c|c|c|}
\hline
Variance & approach & \(N_{A} = 0\) & \(N_{A} = 20\) & \(N_{A} = 30\) & \(N_{A} = 40\) \\
\hline
\multirow{5}{*}{\(2.0 \times 10^{-4}\)} 
 & Worst-case for SIMPLE (for $\beta = 99.9\%$) & \textendash
 & \(2.65 \times 10^{-4}\) & \(1.55 \times 10^{-3}\) & \(1.66 \times 10^{-3}\) \\
\cline{2-6}
 & NRMSE for $\alpha = 70\%$, $\beta = 60\%$ & \(4.98 \times 10^{-5}\) & \(2.56 \times 10^{-4}\) & \(2.56 \times 10^{-4}\) & \(2.56 \times 10^{-4}\) \\
\cline{2-6}
 & NRMSE for $\alpha = 60\%$, $\beta = 60\%$ & \(6.40 \times 10^{-5}\) & \(1.73 \times 10^{-4}\) & \(1.73 \times 10^{-4}\) & \(1.73 \times 10^{-4}\) \\
\cline{2-6}
 & NRMSE for $\alpha = 50\%$, $\beta = 60\%$ & \(6.52 \times 10^{-5}\) & \(1.47 \times 10^{-4}\) & \(1.47 \times 10^{-4}\) & \(1.47 \times 10^{-4}\) \\
\cline{2-6}
 & NRMSE for $\alpha = 40\%$, $\beta = 60\%$ & \(8.99 \times 10^{-5}\) & \(1.18 \times 10^{-4}\) & \(1.18 \times 10^{-4}\) & \(1.18 \times 10^{-4}\) \\
\hline
\multirow{5}{*}{\(2.0 \times 10^{-2}\)} 
 & Worst-case for SIMPLE (for $\beta = 99.9\%$) & \textendash
 & \(2.65 \times 10^{-3}\) & \(1.52 \times 10^{-2}\) & \(1.63 \times 10^{-2}\) \\
\cline{2-6}
 & NRMSE for $\alpha = 70\%$, $\beta = 70\%$ & \(4.02 \times 10^{-4}\) & \(2.60 \times 10^{-3}\)  &\(2.60 \times 10^{-3}\) & \(2.60 \times 10^{-3}\)\\
\cline{2-6}
 & NRMSE for $\alpha = 60\%$, $\beta = 70\%$ &\(4.70 \times 10^{-4}\) & \(1.99 \times 10^{-3}\) & \(1.99 \times 10^{-3}\)& \(1.99 \times 10^{-3}\)\\
\cline{2-6}
 & NRMSE for $\alpha = 50\%$, $\beta = 70\%$ & \(5.50 \times 10^{-4}\) & \(1.43 \times 10^{-3}\) & \(1.43 \times 10^{-3}\) & \(1.43 \times 10^{-3}\)\\
\cline{2-6}
 & NRMSE for $\alpha = 40\%$, $\beta = 70\%$ &\(6.60 \times 10^{-4}\) & \(1.13 \times 10^{-3}\) & \(1.13 \times 10^{-3}\) & \(1.13 \times 10^{-3}\)\\
\hline
\end{tabular}
\vspace{0.1cm}

\textit{Note: The NRMSEs in this table are obtained by using MLE as the fusion approach.}
\end{table*}
\begin{table}[h!]
    \caption{Comparison between NRMSEs for additional processing with $m=150, m=70,000, $ and $m= 300,000$.}
    \centering
    \begin{tabular}{|l|c|}
        \hline
        \textbf{Parameter} & \textbf{Setting or measurement} \\ \hline
        Noise Distribution & Gaussian \\ \hline
        Attack Type  & Non-Random Water-Filling Attack \\ \hline
        Variance & $1 \times 10^{-4}$ \\ \hline
        $\alpha$ & 60\% \\ \hline
        $\beta$ & 80\% \\ \hline
        $N$ & 50 \\ \hline
        $N_{A}$ & 40 \\ \hline
        NRMSE (150 Steps) & $1.86 \times 10^{-4}$ \\ \hline
        NRMSE (70,000 Steps) & $2.80 \times 10^{-4}$ \\ \hline
        NRMSE (300,000 Steps) & $6.00 \times 10^{-4}$ \\ \hline
    \end{tabular}

    \label{tab:simulation_results}
\end{table}



\subsubsection{Water-filling Attack - Worst-case attack on APCC-ADDITIONAL processing}

The previously considered edge attack is the worst-case (causes the most damage) attack for the APCC-SIMPLE processing, but the worst-case attack for the APCC-ADDITIONAL processing is the water-filling attack we discussed previously. In the water-filling attack considered in this subsection, we optimally select the sensors to attack, so we call this a nonrandom attack.  
Table~\ref{ApccAWFNonrandomGasussian} illustrates the APCC-ADDITIONAL processing performance (NRMSE with MLE fusion) for several values of $\alpha$ and $\beta$ for the nonrandom water-filling attack for some cases with Gaussian sensor noise. 
 Note that while the majority of the results in Table~\ref{ApccAWFNonrandomGasussian} are for 
water-filling attacks (worst-case for ADDITIONAL), the first row in Table~\ref{ApccAWFNonrandomGasussian} below the headings is for 
edge attacks (worst-case for SIMPLE) to 
allow simple comparison. 

The results in Table~\ref{ApccAWFNonrandomGasussian} show that, with proper choice of the parameters $\alpha$ and $\beta$, we can find approaches that make the worst-case NRMSE performance of the APCC-ADDITIONAL processing with a large $N_A$  smaller than that of the 
worst-case NRMSE performance of 
APCC-SIMPLE processing 
in the same case and close to GE.  
As Table~\ref{ApccAWFNonrandomGasussian} illustrates, this decrease in worst-case NRMSE performance 
of the APCC-ADDITIONAL processing for cases with a large $N_A$ comes with a small increase in the NRMSE for cases with $N_A=0$ when compared with the APCC-SIMPLE approach in Table~\ref{ApccAEdgeNotrandomAttack} 
(same as GE).
Table~\ref{ApccAWFNonrandomGasussian} shows that 
when the noise distribution is Gaussian with a variance of $1 \times 10^{-4}$, using 
$\alpha=80\%$ and $\beta=80\%$ yields worst-case performance with $N_A=40$ that is within a factor of $2$ of the GE for that case while yielding performance with $N_A=0$ that is within a factor of $1.8$ of the GE for that case.  While this seems to provide good performance at either extreme, other choices of $\alpha$ and $\beta$ 
allow different tradeoffs. 

\begin{figure}[htb]
    \centering
    \includegraphics[width=0.5\textwidth]{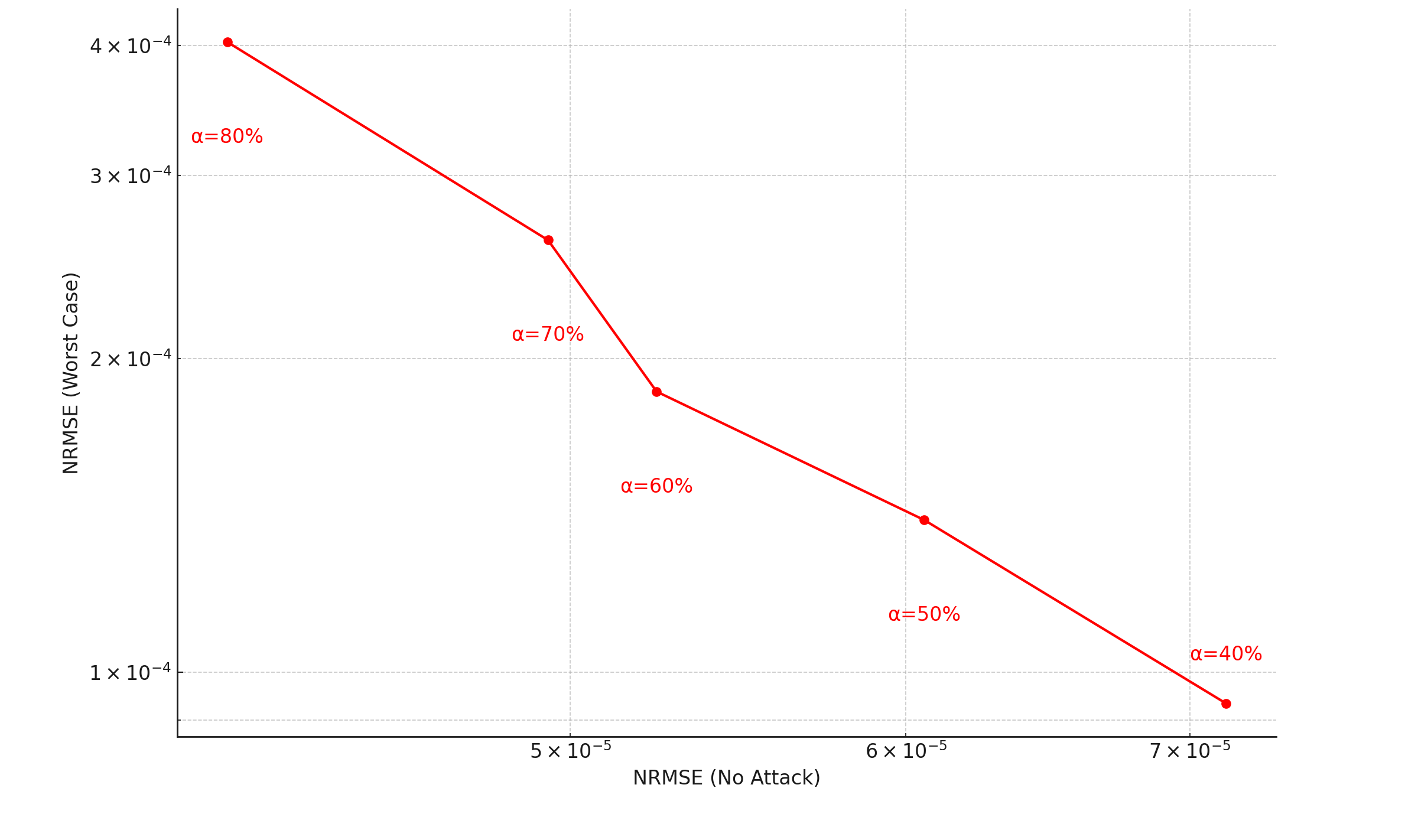}
    \caption{Trade-off between worst-case NRMSE with $N_A = 40$ (Y-axis) versus NRMSE for $N_A = 0$ (X-axis) for a case with Gaussian sensor noise, various $\alpha$, $N = 50, \beta = 80\%$, and noise variance = $1.0 \times 10^{-4}$.}
    \label{fig:curve3}
\end{figure}

\begin{figure}[htb]
    \centering
    \includegraphics[width=0.5\textwidth]{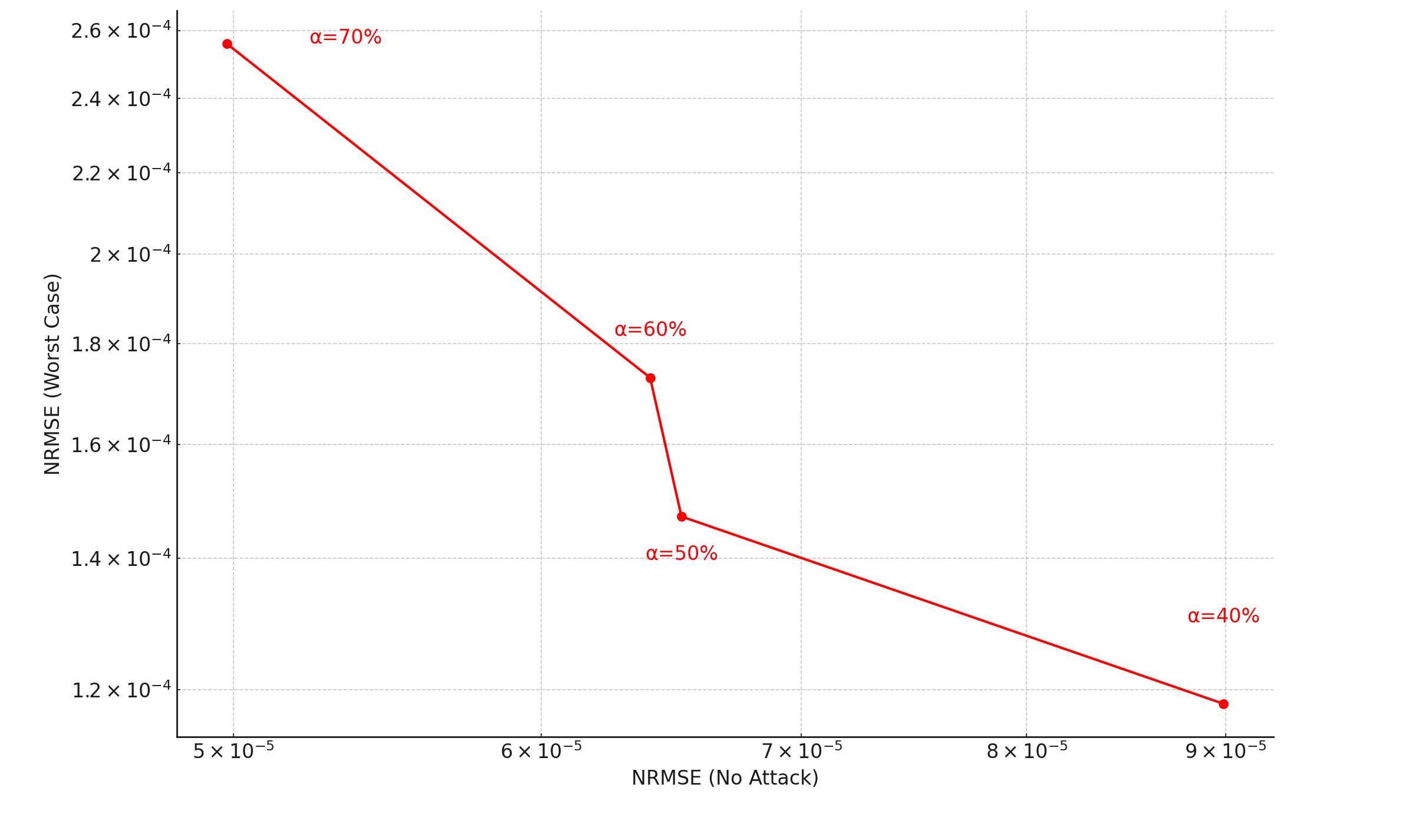}
    \caption{Trade-off between worst-case NRMSE with $N_A = 40$ (Y-axis) versus NRMSE for $N_A = 0$ (X-axis) for a case with Laplacian  sensor noise, various $\alpha$, $N = 50, \beta = 60\%$, and noise variance = $2.0 \times 10^{-4}$.}

    \label{fig:curve4}
\end{figure}

Table~\ref{ApccAWFNonrandomLaplacian60} illustrates similar findings for cases with Laplacian noise. 
Figure~\ref{fig:curve3} illustrates the trade-offs that the different previously considered $(\alpha,\beta)$ approaches can achieve in terms of worst-case NRMSE performance ($y$-axis) for a large $N_A$  versus the $N_A=0$ NRMSE performance ($x$-axis) under Gaussian sensor noise. In addition, Figure~\ref{fig:curve4} shows the trade-offs in worst-case NRMSE performance ($y$-axis) for a large $N_A$  compared to the no-attack NRMSE performance ($x$-axis) when subjected to Laplacian sensor noise.


While the just given results numerically indicate that a sufficient 
decrease in $\alpha$ will 
lead to a decrease in 
the worst-case performance with a large $N_A$, this can be deduced based on the construction of the APCC-ADDITIONAL approach. This follows since reducing $\alpha$ 
reduces the upper bound $U(x)$ which then  eliminates the space between the upper bound and the histogram for all bins  which can support one or more attacks. Thus the bins that have space for attacks to pass through become fewer as $\alpha$ is reduced.  
In fact, 
the most damaging bins of this type will typically 
be heavily impacted 
in this way first so that 
the attacks that pass the protection after $\alpha$ reduction will cause less damage to the estimation/fusion.  It is also clear that increasing $\alpha$ will improve the performance under no attack since fewer sensors will be deemed to be attacked with other things equal. 

It’s important to note that any error in the prediction in 
Fig.~\ref{fAPCCS}  could negatively affect the performance of the proposed APCC approach. 
We omit results showing this since the idea is quite intuitive. 
An accurate prediction approach must be employed. 
One might be concerned that there could be 
an accumulation of prediction errors over time. 
Thus if some attacked  sensors are determined to be unattacked at a given time step,  
these sensors will be used in the prediction for the next time step as described in 
Fig~\ref{fAPCCS}, possibly  increasing the prediction error at the next step. This error could accumulate over time if the attacks always move the prediction in a consistent direction. 
However, 
experimental results show 
if we employ a good machine learning approach which is properly trained then by properly adjusting $\beta$ and $\alpha$ in APCC-ADDITIONAL,  the impact of even powerful optimized attacks  always moving the prediction in a consistent direction 
do not cause large degradation 
over longer trajectories.  Table~\ref{tab:simulation_results} shows results when the noise distribution is Gaussian with a variance of $1 \times 10^{-4}$ for an attacker  launching 40 optimum (non-random water-filling,  consistent direction over time) attacks each time step on a total of 50 sensors. 
Table~\ref{tab:simulation_results}  show that by setting $\alpha$ to 60\% and $\beta$ to 80\%, the APCC-ADDITIONAL approach provides enough protection to counteract even this severe worst-case attack. The simulation results show that the NRMSE reaches $1.86 \times 10^{-4}$ after $m = 150$ time steps. 
{
Even when increasing $m$ to $m = 7.0 \times 10^{4}$ and $m = 3.0 \times 10^{5}$ time steps, the NRMSE increases only slightly, to $2.80 \times 10^{-4}$ and $6.00 \times 10^{-4}$, corresponding to approximately $1.5\times$ and $3.2\times$ the original value, respectively. 
}
These results demonstrate that with proper protection, the  
damage from 
attacks leaking 
through and degrading the  prediction can be made small, even under the worst-case attack, allowing for acceptable performance.

{
\subsection{Comparisons with Subspace-based Sparse Attack Detection~(SSAD) Approach}\label{subsec:recon}


\begin{table}[htb]
\centering
\caption{
SSAD for Gaussian noise with specific shift of $1.0 \times 10^{4}$. See TABLE I for column definitions. $m=150$.}
\label{tab:paper_method_results}
\begin{tabular}{|c|c|c|c|c|}
\hline
$\sigma^2$ & $N$ & $N_{A}$ & $\text{NRMSE SIMPLE}$ & $\text{NRMSE SSAD}$ \\
\hline
$1.0 \times 10^{-4}$ & 50 & 40 & $5.03 \times 10^{-5}$ & $1.22 \times 10^{+2}$ \\
$1.0 \times 10^{-4}$ & 50 & 30 & $3.56 \times 10^{-5}$ & $9.23 \times 10^{+1}$ \\
$1.0 \times 10^{-4}$ & 50 & 10 & $2.52 \times 10^{-5}$ & $3.10 \times 10^{+1}$ \\
$1.0 \times 10^{-4}$ & 20 & 10 & $5.03 \times 10^{-5}$ & $7.87 \times 10^{+1}$ \\
$1.0 \times 10^{-4}$ & 10 & 5  & $7.12 \times 10^{-5}$ & $7.77 \times 10^{+1}$ \\
$1.0 \times 10^{-2}$ & 50 & 40 & $5.03 \times 10^{-4}$ & $1.22 \times 10^{+2}$ \\
$1.0 \times 10^{-2}$ & 50 & 30 & $3.56 \times 10^{-4}$ & $9.23 \times 10^{+1}$ \\
$1.0 \times 10^{-2}$ & 50 & 10 & $2.52 \times 10^{-4}$ & $3.10 \times 10^{+1}$ \\
$1.0 \times 10^{-2}$ & 20 & 10 & $5.03 \times 10^{-4}$ & $7.87 \times 10^{+1}$ \\
$1.0 \times 10^{-2}$ & 10 & 5  & $7.12 \times 10^{-4}$ & $7.77 \times 10^{+1}$ \\
\hline
\end{tabular}
\end{table}

{

In Table~\ref{tab:paper_method_results} for $m=150$, we compare the performance of the subspace-based sparse attack detection approach (SSAD) described in \cite{9992077} with that of our APCC-SIMPLE approach.
In Table~\ref{tab:paper_method_results} 
we always launch 
an attack that 
shifts 
the actual trajectory at attacked sensors by adding 
$1.0 \times 10^{4}$. 
The details of the number of sensors attacked, total number of sensors and the variance of the Gaussian noise added are given in  ($N,N_A,\sigma^2$) in 
Table~\ref{tab:paper_method_results}. 
The particular sensors attacked are chosen randomly. 
Consider the case where 40 out of 50 sensors are under attack and the variance of the Gaussian noise is $1.0 \times 10^{-4}$. The NRMSE of the APCC-SIMPLE approach is $5.03 \times 10^{-5}$, which is close to the NRMSE of GE, while the NRMSE of SSAD is $1.22 \times 10^{+2}$, which is much larger. According to the results in Table~\ref{tab:paper_method_results}, our APCC-SIMPLE approach consistently yields much lower NRMSEs in all the cases tested in the numerical experiment.

To be fair, the attack detection approach in 
\cite{9992077} was shown to  perform well  for 
sparse attacks in 
\cite{9992077}, for which it is designed, but not for the non-sparse attacks. This explains  
Table~\ref{tab:paper_method_results}.  Note that there are many other approaches, as discussed in the literature review, which are designed for cases where less than $k$ sensors are attacked (or similar assumptions), 
However, these methods tend to perform very poorly for other cases (more than $k$ sensors attacked) and all cases are allowed by our attack model.  
}

}

\subsection{Comparison with Data-Reconstruction~(DR) Approach}

\begin{table}[htb]
\centering
\caption{
DR for Gaussian noise with specific shift of $1.0 \times 10^{4}$. See TABLE I for column definitions. $m=150$.}
\label{tab:reconstruction_results}
\begin{tabular}{|c|c|c|c|c|}
\hline
$\sigma^2$ & $N$ & $N_{A}$ & $\text{NRMSE GE}$ & $\text{NRMSE RECON}$ \\
\hline
$1.0 \times 10^{-4}$ & 50 & 40 & $5.03 \times 10^{-5}$ & $1.26 \times 10^{2}$ \\
$1.0 \times 10^{-4}$ & 50 & 30 & $3.56 \times 10^{-5}$ & $9.48 \times 10^{1}$ \\
$1.0 \times 10^{-2}$ & 50 & 40 & $5.03 \times 10^{-4}$ & $1.26 \times 10^{2}$ \\
$1.0 \times 10^{-2}$ & 50 & 30 & $3.56 \times 10^{-4}$ & $9.48 \times 10^{1}$ \\
\hline
\end{tabular}
\end{table}

In Table~\ref{tab:reconstruction_results} for $m=150$, we compare the performance of the data-reconstruction (DR) approach from \cite{yan2025secure} with that of our APCC-SIMPLE method. The total number of sensors \(N\), number of attacked sensors \(N_A\), and the Gaussian noise variance \(\sigma^2\) are listed in Table~\ref{tab:reconstruction_results}, and the attack procedure is described in Subsection~\ref{subsec:recon}. 

Consider the representative case with \(N=50\), \(N_A=40\), and \(\sigma^2=1.0\times 10^{-4}\). The NRMSE of APCC-SIMPLE is \(5.03\times 10^{-5}\), which is close to the GE, whereas the NRMSE of the DR method is \(1.26\times 10^{2}\), several orders of magnitude larger. Across all tested cases, APCC-SIMPLE yields substantially lower NRMSE and is always close to the GE. 

To be fair, the attack detection approach in  \cite{yan2025secure} was shown
to perform well for certain assumptions on the 
number of attacked sensors.  These assumptions are not true for all attacks in our attack model, which explains the bad behavior. 


{ 
\subsection{Comparisons with EDAD Approaches}

As previously discussed, one can always substitute any valid trajectory for another valid trajectory and the substituted trajectory will pass EDAD since it is valid. 
Such substitutions are possible, as long as the shape of the trajectory after attack is similar to what 
is expected based on the training data. 
There are always cases where these substitutions can lead to very large errors as we try to explain here.   
Let's focus on a sensor measuring 
position since we focus on this in the numerical examples.  
Let's also focus on the case where the actual  trajectory $ y_1,y_2,\ldots,y_m $ in 
the computation of 
(\ref{NRMSE}) 
satisfies $ {\frac{1}{m}\sum_{j=1}^{m}|y_j|} = 1 $ so we can show bad performance for EDAD for some cases in a simple way.   
Thus, an easy way to change the position trajectory 
to another valid 
trajectory 
$ \hat{y}_1,\hat{y}_2,\ldots,\hat{y}_m $, with similar shape,  
is to shift all the elements of the position trajectory by some constant $h$, 
thus $\hat{y}_j = y_j + h $ for all $1 \leq j \leq m $.   
The large values of $h$ that are possible in typical applications lead to large NRMSEs in 
(\ref{NRMSE}).  For example, a shift so that  $\hat{y}_j = y_j + h$ for all $1 \leq j \leq m $, yields $NRMSE=h$.
Thus if $h$ is unbounded, so is $NRMSE$. 
Even if you limit the trajectories to come from a specified area, 
the errors can be  very large. One can estimate the diameter of Bethlehem PA as approximately 3970 meters.  Thus if we set $h=3970$, then $NRMSE = 3970$
which is much larger than even the large errors in Table~\ref{tab:paper_method_results}. The point is that 
attacks which pass EDAD can lead to extremely large errors so one can't employ this approach since you can't typically tolerate these large errors. 

Note that both of our approaches build on the simple approach. The simple approach limits the size by which the attack can change the sensor data. If the sensor data is changed by too large an amount, due to the confidence interval comparison in 
Fig \ref{fAPCCS}, the attack will be blocked. On the other hand, large attacks can  pass EDAD if they follow valid trajectories. 
To pass our approaches, the attack can't be too different from the "typical unattacked sensor prediction" in 
Fig \ref{fAPCCS}. 
How large an attack is allowed is controlled by $\beta$.   The size allowed is very small in all the cases we considered, on the order of $\beta \%$ of the sensor noise in the training data. 
Thus our approaches limit the size of attacks, while EDAD does not. 
}

\newpage 

\section{Conclusions And Future Work}
\label{sectcon}


{This paper developed 
methods that  accurately
identify/eliminate just the problematic attacked sensor data
(keep the rest) presented to a sequence estimation/regression algorithm for all possible attacks under our attack model. 
} The developed approaches were shown to protect a  sequence  estimation/regression algorithm designed for
unattacked sensor data by  allowing such an algorithm to operate using the deemed to be unattacked data. The developed approaches
employ only unattacked training data to  mitigate all 
attacks allowed { under our attack model} where the
attacked sensors and attacks can change for each time step,
the attacker has complete control of after-attack sensor data,
and the protection system has no prior knowledge of 
how many sensors are
attacked and which
sensors are more likely to be attacked. 

We first proposed a  simple protection
approach for attacks not endowed with knowledge of the
details of our protection approach, but later we proposed additional processing for attackers with knowledge.  This  allows a lower complexity approach to be employed if the attacker does not have detailed knowledge of the protection approach. Experimental results were provided  which demonstrated good performance for both the
simple and additional processing. Our simple method is shown
to achieve estimation performance which is indistinguishable (to two decimal places) when compared to that of the GE (which knows which sensors were attacked) 
in cases 
tested, which assume the attacker does not have knowledge
of the protection scheme. For cases
where the attacker has knowledge of the protection approach, numerical results show that 
the additional processing 
can be configured, by proper parameter choice, so that  the worst-case degradation under a large
number of sensors attacked can be made significantly smaller
than the worst-case degradation under the simple  processing, and close to GE, 
for the same number of attacked sensors with just a slight
degradation under no attacks. Guarding against the worst-case is extremely important and if this performance is acceptable, then performance will always be acceptable. 

We were able to mathematically describe the worst-case attacks for our simple and additional 
processing approaches, which allowed us to describe how
to calculate the worst-case performance. 
The mathematical description of our worst-case attacks allowed us to show that 
proper choice of the additional processing parameters imply
the worst-case degradation under the additional processing and
a large number of sensors attacked can be made 
smaller, even for cases we did not test numerically.  
Since only a limited number of different cases can be numerically tested, this is important.  We
hope other researchers  will employ similar ideas in their future work since these ideas seem very powerful. 
Directly calculating the worst-case performance,
by knowing the worst-case attack is extremely efficient in
reducing computations, which is especially important if you
want to try many parameter settings. It avoids trying many different
attacks, an approach others take. Such an approach would be difficult
due to extremely high complexity and one would never get
the actual worst-case attack performance.

We recognize that we have only taken the first steps in a new direction.  There are still many more details that should be further investigated in the future and we list some of these in the rest of this section. 
The numerical results provided are extensive but are limited as any numerical results would be.  It would be of interest to expand the cases tested in many ways.  
For example, in the presented numerical results, we focus solely on scalar trajectories, but in other investigations, not reported here, we tested the proposed algorithm on some higher-dimensional cases as well.  We found the algorithm provides similarly good performance, compared to the cases reported here.  Regardless, further testing would be desirable.  
As another example, we tested for a few specific statistical models, but we could expand the statistical models for the sensor observations to include many more models.   It would also be of great interest to test using measured data.  While we did not find suitable measured data to date, there are discussions of  future initiatives which could provide this data in the future.  

It would be of interest to further study methods to ensure good predictions where predictions are used in our approaches and maybe to monitor these predictions to make sure nothing has gone wrong.  One issue 
of concern is to make sure problematic attacks 
do not leak through our protection to impact our predictions. 
Fortunately,  we have found that this can be ensured even under very aggressive attacks if the APCC-ADDITIONAL approach employs $\alpha$ and $\beta$ which provide the needed level of protection.  On the other hand, it would be of great interest to carefully study methods which might further enhance performance {
in } these cases. This seems possible.  {  Finally, it would be useful to study attacks on the training data, which we did not consider here. }



In closing, it should be noted that our approach can be considered one level of protection which can be combined with other levels to enhance overall protection.  It would be interesting to understand the gains of combining multiple approaches in this manner.

\end{document}